\documentclass[preprint,12pt]{elsarticle}

\usepackage{amssymb}
\usepackage{graphicx}
\usepackage{caption}
\usepackage{subcaption}
\usepackage{xcolor}
\usepackage{amsmath}
\usepackage{multirow}
\usepackage{makecell}
\usepackage[figuresright]{rotating}
\usepackage[linesnumbered,ruled,vlined]{algorithm2e}
\newcommand{\methodname}{{\tt{FLEEG}}}

\journal{Nuclear Physics B}

\begin{document}

\begin{frontmatter}



\title{Aggregating Intrinsic Information to Enhance BCI Performance through Federated Learning}


\author[label1]{Rui Liu}
\author[label1]{Yuanyuan Chen}
\author[label1]{Anran Li}
\author[label1]{Yi Ding}
\author[label1]{Han Yu}
\author[label1]{Cuntai Guan\corref{cor1}}
\affiliation[label1]{organization={School of Computer Science and Engineering, Nanyang Technological University}, 
            addressline={50 Nanyang Ave}, 
            postcode={639798}, 
            country={Singapore}}
\cortext[cor1]{Corresponding author}

\begin{abstract}
Insufficient data is a long-standing challenge for Brain-Computer Interface (BCI) to build a high-performance deep learning model. Though numerous research groups and institutes collect a multitude of EEG datasets for the same BCI task, sharing EEG data from multiple sites is still challenging due to the heterogeneity of devices. The significance of this challenge cannot be overstated, given the critical role of data diversity in fostering model robustness. However, existing works rarely discuss this issue, predominantly centering their attention on model training within a single dataset, often in the context of inter-subject or inter-session settings.
In this work, we propose a hierarchical personalized Federated Learning EEG decoding (\methodname{}) framework to surmount this challenge. This innovative framework heralds a new learning paradigm for BCI, enabling datasets with disparate data formats to collaborate in the model training process. Each client is assigned a specific dataset and trains a hierarchical personalized model to manage diverse data formats and facilitate information exchange. Meanwhile, the server coordinates the training procedure to harness knowledge gleaned from all datasets, thus elevating overall performance.
The framework has been evaluated in Motor Imagery (MI) classification with nine EEG datasets collected by different devices but implementing the same MI task.
Results demonstrate that the proposed frame can boost classification performance up to 16.7\% by enabling knowledge sharing between multiple datasets, especially for smaller datasets.
Visualization results also indicate that the proposed framework can empower the local models to put a stable focus on task-related areas, yielding better performance.
To the best of our knowledge, this is the first end-to-end solution to address this important challenge.

\end{abstract}




\end{frontmatter}


\section{INTRODUCTION}
\label{sec:intro}

Brain-computer interface (BCI) is a crucial technology that establishes a connection between the human brain and external devices, which has clinical and non-clinical applications in many areas, such as movement capability recovery and assistance \cite{mane2020bci}, cognitive health \cite{lee2013brain}, and entertainment \cite{nijholt2022brain}.
Electroencephalogram (EEG) is one of the most commonly used signals in BCI to decode brain activities.
In recent years, deep learning algorithms have been employed in EEG decoding and classification tasks. 
Since the success of deep learning algorithms is largely attributed to the availability of large amounts of data, collecting enough data is important for EEG decoding.
However, EEG data collection from humans can be challenging and costly.
The physiological limitations of subjects limit the number of samples collected from one person.
And the significant cost and complicated usage of the devices limit the number of participants in one dataset. Thus, most datasets are of middle size, collected from dozens of subjects with hundreds of samples per subject at most.

Existing works enhance EEG decoding model performance mainly by sharing knowledge between subjects or sessions within one single dataset \cite{wan2021review,wei2021inter}. However, 
there are some datasets designed with the same task and protocol.
Drawing upon the widely acknowledged principle that larger training datasets tend to yield enhanced classification outcomes in deep learning approaches, it is anticipated that advancements in model performance could be realized through the amalgamation of knowledge from these datasets as a larger virtual training set.
Unfortunately, EEG data collected by various devices have heterogeneous formats, in terms of the number and location of EEG channels, sampling rates, and amplifiers.
This device heterogeneity problem prevents knowledge sharing among datasets. 
Few works made initial attempts to solve this problem by dropping channels or padding with zeros, which may lose information or add noise to the data \cite{gu2022multi,saeed2021learning,bakas2022team}. 
This important issue remains unresolved.

Federated learning (FL) is an emerging collaborative machine learning paradigm to train models across multiple data owners and enables local models to benefit from each other while keeping local data private \cite{kairouz2021advances}.
FL has been applied in many medical and healthcare applications, such as disease prediction \cite{peng2022fedni}, and brain template estimation \cite{bayram2021federated}, to facilitate cross-silo cooperation with privacy protection.
Personalized Federated Learning (PFL) \cite{tan2022towards} is a branch of FL dealing with the heterogeneity issue across clients, including the heterogeneity in data distributions, network structures, and data formats. It inspires us to design the proposed framework to solve the device heterogeneity issue in the EEG decoding application.

In this work, we provide a new solution to enlarge the training set by including more related datasets. We design a hierarchical personalized \underline{F}ederated \underline{L}earning \underline{EEG} decoding (\methodname{}) framework to solve the device heterogeneity issue. 
The framework makes use of the PFL architecture to facilitate cooperative training among multiple device-heterogeneous EEG datasets. 
It consists of a server and several clients. The server orchestrates local models' training in clients to obtain the optimal personalized models for each dataset assigned to the clients.
The personalized model compromises a local module followed by a global one. 
The local module is responsible for extracting features with the same formats from device-heterogeneous datasets, while the global module transfers the knowledge between datasets to improve model performance.
We evaluated the proposed framework with nine real Motor Imagery (MI) EEG datasets collected by multiple institutes.
It improves the model performances on most datasets compared to independent training, especially for smaller datasets with an improvement of up to 16.7\%.
This framework provides a new general learning paradigm for the BCI community to train higher-performance models with multiple datasets, instead of one.

In summary, to the best of our knowledge, this is the first endeavor to tackle the device heterogeneity issue among multiple EEG datasets with a federated-learning-based end-to-end solution for training deep learning models. 
We briefly summarize the contributions of our work as follows:
\begin{itemize}
    \item To obtain a higher-performance model, we provide a new learning paradigm for the BCI community to train the EEG decoding models with an enlarged training set consisting of multiple datasets, instead of one dataset.
    \item To solve the device heterogeneity issue, we propose a federated-learning-based framework, named \methodname{}, enabling knowledge sharing between datasets during the model training process. Each client is assigned a dataset and trains a hierarchical personalized  model, consisting of a local module to align data formats and a global module to transfer knowledge between datasets. 
    \item To validate the performance of the proposed framework, we evaluate the proposed framework on nine real EEG MI datasets. The results demonstrate remarkable improvements in the performance of local models across the majority of datasets, especially for small datasets with an improvement of up to 16.7\%. We further analyze the factors for the improvements and provide visualized interpretation. 
\end{itemize}

The paper is structured in the following way: Section \ref{sec:related_works} covers related works, Section \ref{sec:methods} explains the proposed framework in detail, and Section \ref{sec:evaluations} describes the experiment setups and reports the experiment results with analysis. The paper concludes in Section \ref{sec:conclusion}.

\section{RELATED WORKS}
\label{sec:related_works}

\subsection{Device-heterogeneity issue in EEG}
\label{sec:heterogeneity-issue}
The issue of device-heterogeneity in EEG applications has received limited attention in existing studies. Most of the research has primarily focused on transferring knowledge across subjects or sessions within one dataset \cite{wan2021review,zhang2020application,wei2021inter}.  
Only a few works have made attempts to address the heterogeneity across datasets in EEG applications. Some works \cite{gu2022multi,saeed2021learning,bakas2022team,xu2020cross,kuang2021cross,cuibenchmarking} make the data formats consistent by either deleting channels or padding with zeros. However, these methods can introduce disturbances into the EEG signal, either by losing valuable information or adding noise.
Other researchers have explored the channel mapping methods as a separate feature extractor during pre-processing on heterogeneous datasets \cite{kostas2021bendr}. Nevertheless, since this feature extractor is not integrated into the end-to-end training process, it lacks the flexibility to adapt to each dataset and may introduce noise into the analysis.

\subsection{Federated learning}
Federated learning (FL) is a collaborative machine learning paradigm to train models across distributed data owners and enables local models to benefit from each other. It allows privacy-preserving, especially for applications that use sensitive or personal data.
In FL, data owners with local data can be referred to as \textit{clients} if they are coordinated by a central entity referred to as the \textit{server}. 

Since the data is isolated in the clients, personalized federated learning has been developed to solve the heterogeneity issues across clients \cite{tan2022towards,gao2022survey}.
According to the heterogeneity types, existing in data distribution, data format, local model structure, computation ability and etc, various solutions have been proposed.
For the data format heterogeneity issue, some works \cite{feng2022semi,bica2022transfer,liang2020think} design local encoders to project the data to a common space first and then transfer knowledge between clients based on the aligned features. 

Federated learning algorithms have been applied to some BCI applications recently.
\cite{hang2023fedeeg} applies Federated Learning to EEG decoding with the cross-subject task. \cite{ju2020federated,hu2021cross} borrow the manifold learning methods in transfer learning and apply them in the federated setting on the cross-subject and cross-session tasks.
These works only focus on the data distribution heterogeneity issue limited to one dataset.
Besides, privacy protection between subjects in one dataset is not a critical issue. EEG data is not as intuitive as images. People can understand the information in the image when they see it, but EEG information needs a professional device to collect and methods to decode.
Thus, the ability to interpret and protect EEG data privacy is limited exclusively to research institutes rather than individuals.
Thus, EEG data privacy protection for research institutes is more practical than the individuals.

\section{PROPOSED METHODS}
\label{sec:methods}

Leveraging multiple device-heterogeneous datasets presents a potential solution for expanding the training set for high-performance model training. However, effectively utilizing such datasets remains an unsolved but critical problem. In this section, we first describe this problem and then present our proposed solution.

\subsection{Problem description}
As illustrated in Section \ref{sec:heterogeneity-issue} that the device-heterogeneity issue in EEG model training leads to the small amount of available data and the low test accuracy of the trained model. 
There are two kinds of entities involved: a server $\mathcal{S}$ and $K$ distributed clients (\emph{i.e.},  EEG data collection devices). Each client possesses a dataset $D_k=\{x_k, y_k\}, k=\{1, 2, \cdots, K\}$, where $x_k\in \mathbb{R}^{C_k\times T_k}$ represents EEG recordings and $y_k\in \{0, 1\}$ indicates the corresponding labels. $C_k$ and $T_k$ represent the number of channels and the number of time steps, respectively. Each dataset $D_k$ has $S_{k}$ subjects with $N_k$ trails or samples collected per subject. 
The goal is to make full advantage of  data from various devices to train high-performance models. 

\subsection{Overall framework}

\begin{figure}[t]
  \centering
  \includegraphics[width=\columnwidth]{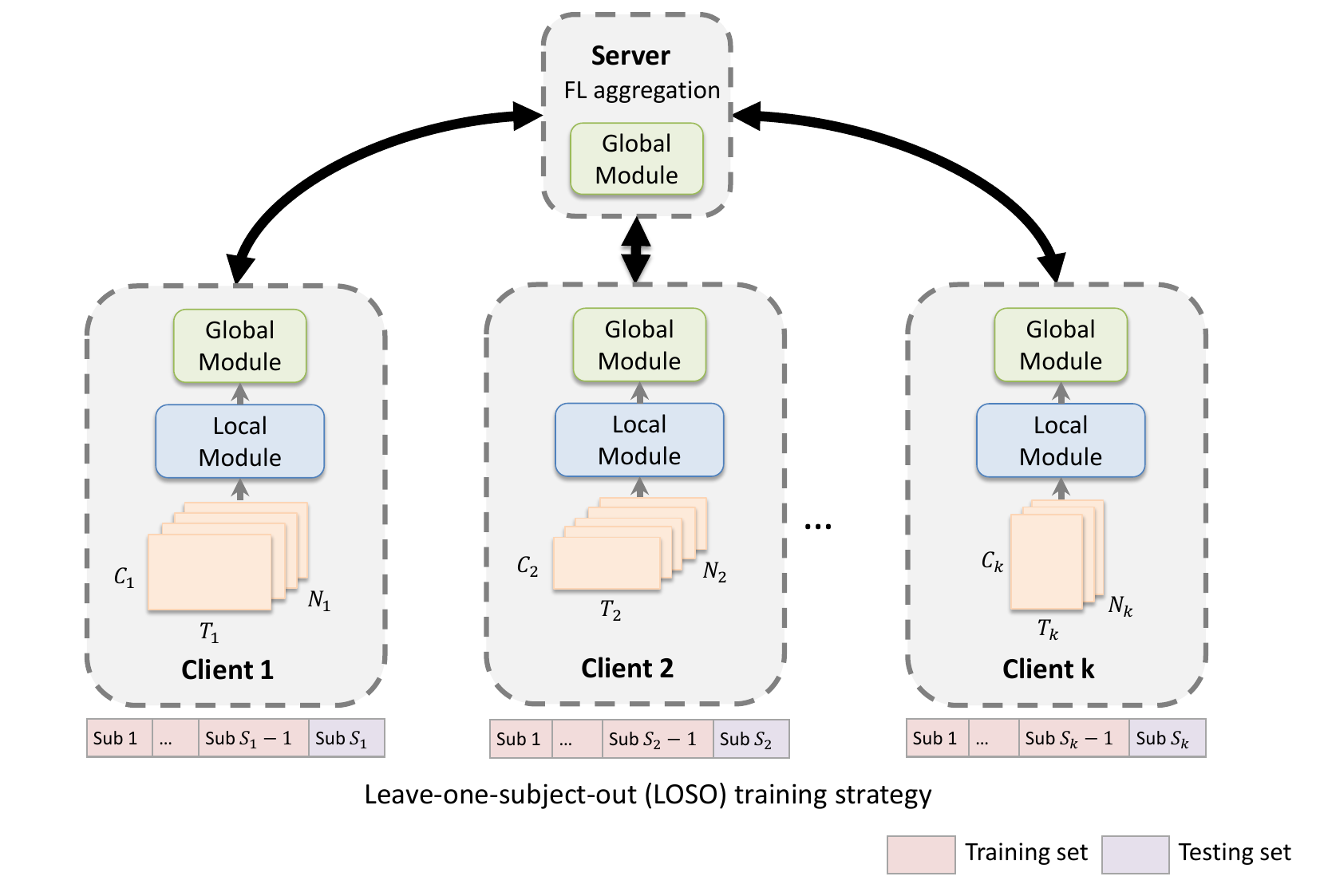}
  \caption{The overview of the proposed hierarchical personalized federated learning framework. Each client is assigned one dataset with various formats defined by $C_k$, $T_k$, and $N_k, k=1,\dots,K$. Clients train their local personalized model, including a local module and a global module, for the classification task. The server manages the cooperation between clients. Each dataset applies the LOSO strategy simultaneously.}
  \label{fig:FLEEG}
\end{figure}

To utilize the device-heterogeneous datasets, we propose a personalized federated learning EEG decoding framework, named \methodname{}.
Following the classical FL framework, the proposed framework consists of one central server and several clients, as illustrated in Figure \ref{fig:FLEEG}.
Each client is assigned one dataset and processes the dataset with its local personalized model for the classification task.
The server manages the cooperation between clients.
Note that the "\textit{personalized model}" in this work indicates the specific network structures designed for the assigned datasets, instead of models trained for different subjects. 

The personalized model in the client consists of a local module and a global module.
The local module acts as a feature encoder to extract the embedding features from EEG data and map them into a unified format across clients. 
The global module is designed to transfer knowledge between clients by communicating the model weights of global modules in all clients via the FL aggregation in the server.
The proposed framework makes each client train its personalized model not only using its own data but also employing knowledge transferred from other datasets, which tremendously enlarges its training set to get better performance.
Next, we introduce the detailed designs of the proposed framework.

\subsection{Personalized model in the clients}
\label{ssec:client_design}

\begin{figure}[t]
  \centering
  \includegraphics[width=\columnwidth]{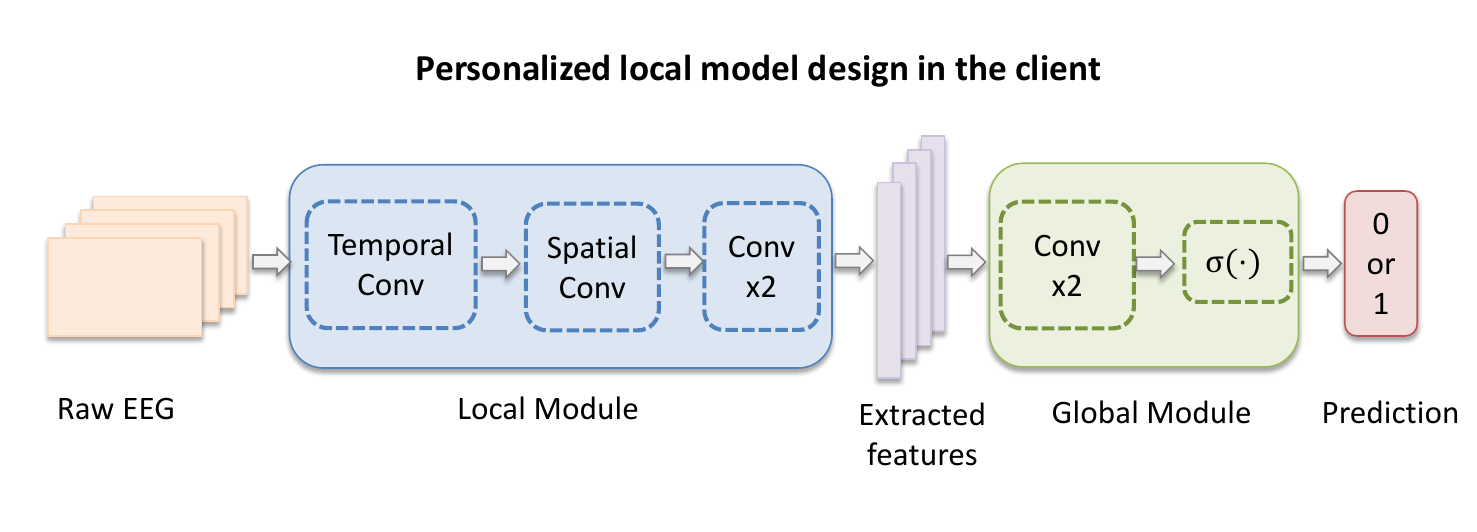}
  \caption{A local personalized model with local and global modules in one client.}
  \label{fig:FLEEG_personalized_model}
\end{figure}

Clients train personalized local models on their corresponding datasets. 
As illustrated in Figure \ref{fig:FLEEG_personalized_model}, the input EEG is first processed by the local module to extract embedding features. Then, the extracted features are sent to the global module to get the prediction as the output.
Inspired by the DeepConvNets (DCN) \cite{schirrmeister2017deep}, we design the local module with a convolution-based temporal filter and a convolution-based spatial filter followed by two standard convolution-max-pooling layers to extract spatial-temporal information from EEG data, refer to the "\textit{Local Module}" part in Figure \ref{fig:FLEEG_personalized_model}.
To match the dataset formats, the network structure design of the local module is personalized to its dataset. 
By setting a suitable kernel size based on the format of input data, the extracted features from heterogeneous datasets can be unified. The detailed settings will be introduced in section \ref{ssec:local_module}.
The global module is designed with one standard convolution-max-pooling layer for high-level feature extraction, followed by a convolution-softmax layer for final classification, refer to the "\textit{Global Module}" part in Figure \ref{fig:FLEEG_personalized_model}.

Since we evaluate the proposed algorithm on a classification task, the local model training is guided by the cross-entropy loss $\mathcal{L}_k(\theta^g,\theta^l_k)$, which is defined as follows:
\begin{equation}
\begin{split}
    \mathcal{L}_k(\theta^g_k,\theta^l_k) = -&\sum_{i} \Bigl[y_ilog\Bigl(g\left(l_k(\textbf{x}_i,\theta^l_k),\theta^g_k\right)\Bigr)\\
    & + (1-y_i)log\Bigl(1-g\left(l_k(\textbf{x}_i,\theta^l_k),\theta^g_k\right)\Bigr)
    \Bigr]
\end{split}
\end{equation}
where $l_k(\cdot,\theta^l_k)$ and $g(\cdot, \theta^g_k)$ describe the local module with its model weights $\theta_k^l$ and the global module with corresponding model weights $\theta^g_k$, respectively.
$\textbf{x}_i$ and $y_i$ represents the $i$-th sample and its label in the dataset.

\subsection{FL aggregation in the server}

Inspired by \textit{FedAvg} \cite{mcmahan2017communication}, the entire framework is trained with the following overall loss function:
\begin{equation}
\mathcal{L}(\theta^g,\theta^l) = \sum_{k=1}^K\frac{N_k}{N} \cdot \mathcal{L}_k(\theta^g_k,\theta^l_k)
\label{eq:FLEEG}
\end{equation}
where $N_k$ denotes the number of samples in client $k$ and $N$ represents the total number of samples in all clients. $\mathcal{L}_k(\theta^g,\theta^l_c)$ is the local loss function in client $k$ as illustrated in section \ref{ssec:client_design}.

The training procedure of the proposed framework follows a classical FL system training process.  
Once the local personalized model finishes local training for the current round, the model weights in the global modules are uploaded to the server for the FL aggregation to transfer knowledge between datasets.
The server updates the model weights of global modules in the server as below:
\begin{equation}
\theta^g = \sum_{k=1}^K\frac{N_k}{N}\cdot \theta^g_k
\label{eq:FedAvg_server}
\end{equation}
After the global module in the server is updated, the server distributes the updated model parameters to the clients. Combined with their local modules, clients continue to train the personalized model with their datasets for the next round. The training process stops until the whole system converges. 
The overall training process of our framework is presented in Algorithm \ref{alg:FLEEG}.

\SetKwInOut{SE}{Server executes}
\SetKwInOut{CU}{ClientUpdate($k,\theta^{g_t}$)}
\SetKwFor{ForAll}{for}{in parallel do}{end}
\begin{algorithm}[hbt!]
\caption{\methodname{}}\label{alg:FLEEG}
\KwIn{local training data from $K$ clients, 
the number of round $R$, 
the number of local epochs $E$, 
the learning rate $\eta$,
the minibatch size $B$}
\SE\\
initialize global modules with weights $\theta^g$; \\
initialize $K$ local modules with weights $\theta^l_k$\;
    \For{each round $t=1,\dots R$}{
        \ForAll{each client $k$}{
            $\theta^{g^{(t+1)}}_k \leftarrow ClientUpdate(k,\theta^{g^{(t)}})$;
        }
        $\theta^{g^{(t+1)}} = \sum_{k=1}^K\frac{N_k}{N} \cdot\theta^{g^{(t+1)}}_k$ \tcp*{aggregate updates}
    }
\CU\\
$B_k$ $\leftarrow$ Split local data into batches of size $B$\;
\For{each local epoch e = 1,\dots,E}{
    \For{batch $b_k \in B_k$}{$\textbf{h}=g\Bigl(l_k(\textbf{x},\theta_k^l),\theta_k^l\Bigr)$\ \tcp*{inference step}
        $\theta_k^l \leftarrow \theta^l_k - \eta \cdot \nabla_{\theta_k^l}\mathcal{L}_k(\theta^g_k,\theta^l_k)$ \tcp*{update local module}
        $\theta_k^g \leftarrow \theta^g_k - \eta \cdot \nabla_{\theta_k^g}\mathcal{L}_k(\theta^g_k,\theta^l_k)$ \tcp*{update global module}
    }
}
\end{algorithm}

\section{Evaluations}
\label{sec:evaluations}
In this section, we first evaluate the effectiveness of \methodname{} by comparing it against the baselines. Then, we provide analysis of how \methodname{} learns robust FL models across heterogeneous EGG data.

\subsection{Experiment Settings}
\label{sec:setups}
We evaluate the proposed FL framework using nine EEG MI datasets. We begin by introducing the experiment settings, including the selection of datasets, network structure settings, evaluation strategies, and the baseline model used for comparison.

\subsubsection{Datasets}
\label{ssec:datasets}
The objective of this work is to transfer knowledge across heterogeneous datasets with the same task but different data formats, including the number of subjects, channels, and sampling frequencies. 
According to this assumption, we select nine public EEG datasets for this study: 
the Korea University (KU) MI dataset \cite{lee2019eeg}, the Shanghai University (SHU) MI dataset \cite{ma2022large}, the Shin2017A dataset \cite{shin2016open}, the BCI-IV-2a dataset \cite{tangermann2012review}, the Weibo2014 \cite{yi2014evaluation}, the MunichMI \cite{grosse2009beamforming}, the High-Gamma Dataset (HGD) \cite{schirrmeister2017deep}, the Cho2017 \cite{cho2017eeg}, and the Murat2018 \cite{kaya2018large} dataset.
These datasets all focus on the hands' motor imagery task to classify subjects' imagery movements of their hands. All of them contain the left-hand and right-hand motor imagery classes.
The statistical information of these datasets is presented in Table \ref{tab:dataset_info} with the number of subjects, the number of trials per subject, the total amount of trials, the number of channels, and sampling frequencies.
It should be noted that some datasets have more than two classes (e.g. the BCI-IV-2a dataset has four categories including left hand, right hand, feet, and tongue), but this work only uses data related to the left and right hand.

\begin{table}[hbt!]
\caption{Statistic information of the four MI EEG datasets}
\label{table_example}
\begin{center}
\resizebox*{1\linewidth}{!}{
\begin{tabular}{c|c|c|c|c|c|c}
\hline
No. & Dataset & \#Subjects & \#Trials/sub & \#Trials &\#Channels  & $f$ (Hz) \\
\hline
1 & KU & 54 & 400 & 21600 & 62 & 1000 \\
\hline
2 & SHU & 25 & 500 & 12500 & 32 & 250 \\
\hline
3 & Shin2017A & 29 & 60 & 1740 & 30 & 1000 \\
\hline
4 & BCI-IV-2a & 9 & 288 & 2592 & 22 & 250 \\
\hline
5 & Weibo2014 & 10 & 158 & 1580 & 60 & 200 \\
\hline
6 & MunichMI & 10 & 300 & 3000 & 128 & 250\\
\hline
7 & High-Gamma Dataset(HGD) & 14 & 482 & 6742 & 128& 500\\
\hline
8 & Cho2017 & 52 & 190 & 9880 & 64 &  512 \\
\hline
9 & Murat2018 & 11 & 1593 &17515 & 22 & 200 \\
\hline
\end{tabular}
}
\end{center}
\label{tab:dataset_info}
\end{table}

The EEG data is band-pass filtered between 0.3Hz and 40Hz. 
To save the usage of RAM, we downsample the KU dataset from 1000Hz to 250Hz and the Shin2017A dataset from 1000Hz to 200Hz. 

\subsubsection{Local models}
\label{ssec:local_module}

As illustrated in the section \ref{ssec:client_design}, we employed the DCN model as the backbone in the clients of the proposed framework.
The network structures of the local modules were designed to accommodate different data formats, as outlined in Table \ref{tab:local_module_network_settings}. On the other hand, the model designs of the global modules remained consistent across all clients and the server, referring to Table \ref{tab:global_module_network_settings}.

\subsubsection{Evaluation settings}
We adopt a subject-independent setting for the cross-dataset training task and utilize the leave-one-subject-out (LOSO) strategy for evaluation. LOSO leaves one subject as the test set. The rest subjects are partitioned into a training set and a validation set to train one model. Once the model is well-trained, it is tested on the left subject to obtain classification accuracy. The overall performance is determined by averaging these accuracy values across all subjects.

Since our evaluation involves nine datasets comprising a total of 214 subjects, strictly following the LOSO approach would be extremely time-consuming. Consequently, we propose an approximate version of the LOSO evaluation strategy for the federated learning framework to enhance training efficiency. Each dataset applies the LOSO methodology independently but simultaneously, as illustrated in Figure \ref{fig:FLEEG}.
Compared to the strict LOSO approach, the modified version leaves out the data of nine subjects - one from each dataset - as nine test sets for nine datasets correspondingly. However, due to variations in the number of subjects across datasets, subjects from datasets with fewer individuals will undergo the LOSO multiple times. This repetition is aimed at facilitating the training for datasets with more subjects. Ultimately, the final classification results for these repeated subjects are obtained by averaging the values.

In the experiment, the partition of the training set and the validation set follows a trial-wise way with a ratio of 9:1. 
We set the maximum training round $R$ to 250 and local epoch $E$ to 1. Once the model training finishes, the model with the smallest validation loss is selected as the best model and applied to the test set to get classification accuracy for the subject. 
The final result for the target client is the averaged classification accuracy of all subjects.
Batch sizes $B$ and learning rates $\eta$ for different datasets are set differently, please refer to the Tabel \ref{tab:lr_bs}.

\begin{table}[h!]\small
\centering
\begin{tabular}{ccc}
\hline
Dataset   & Learning rate & Batch size \\ \hline
KU        & 0.01          & 512          \\
SHU       & 0.005         & 512           \\
Shin2017A & 0.005         & 512           \\
BCI-IV-2a & 0.01          & 512           \\
Weibo2014 & 0.005         & 512           \\
MunichMI  & 0.01          & 128           \\
HGD       & 0.01          & 128           \\
Cho2017   & 0.01          & 512           \\
Murat2018 & 0.01          & 512           \\ \hline
\end{tabular}
\caption{The learning rate and batch sizes used in the training of each dataset.}
\label{tab:lr_bs}
\end{table}

\subsubsection{Comparison baselines}
Given that no prior research has addressed this issue in BCI application, we compare the proposed algorithm with a baseline approach that trains the models independently using a single dataset. To ensure a fair comparison, we maintain consistency between the network structures, LOSO training strategy, maximum training rounds, learning rates, and batch sizes for both the proposed algorithm and the baseline approach.
The network structures in the baseline setting are the same as the local personalized model structures (including the local module and the global module) in the clients for the corresponding datasets in the proposed framework.

\subsection{Experiment Results}
\label{sec:results}
In this section, we present the results of the proposed framework on nine distinct datasets. We proceed to examine the factors influencing the performance of \methodname{}, followed by a visual representation illustrating the improvements achieved by the proposed framework.


\subsubsection{\methodname{} improves performance}

\begin{figure}[h]
  \centering
  \includegraphics[width=\columnwidth]{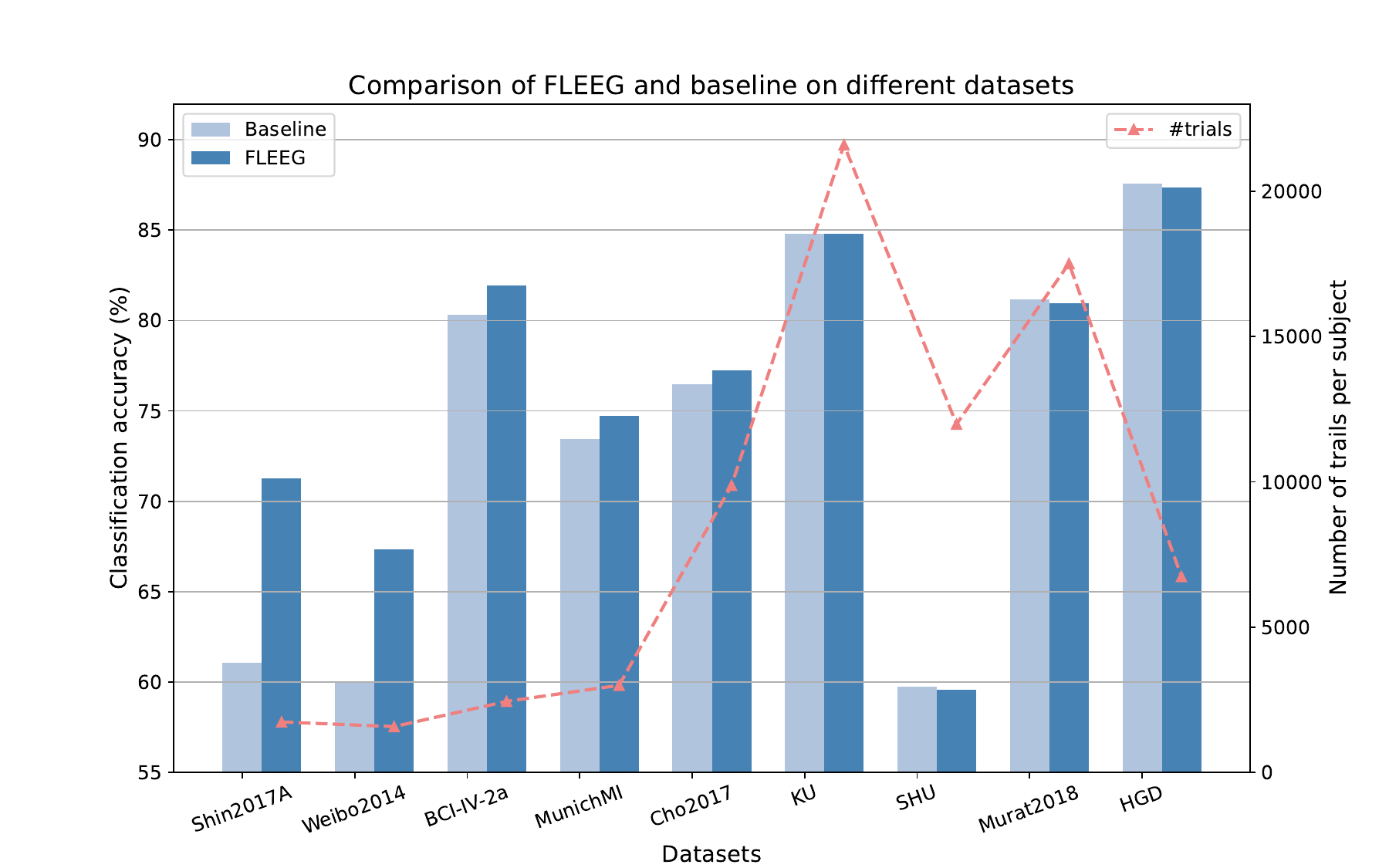}
  \caption{The proposed algorithm and the baseline were compared on nine MI EEG datasets, with the classification accuracy results presented in ascending order based on their respective improvements. The number of trials for each dataset is also shown. The proposed framework effectively supports datasets, especially those that are smaller.}
  \label{fig:results_NumSub_increasement}
\end{figure}

We conducted a comparison between the proposed algorithm and the baseline. Figure \ref{fig:results_NumSub_increasement} showcases the averaged classification accuracies of \methodname{} and the baseline on each dataset. The results of datasets are arranged in ascending order based on their respective improvements.





According to Figure \ref{fig:results_NumSub_increasement}, 
the proposed framework has resulted in significant improvements in the classification accuracy of various datasets. Specifically, the Shin2017A, Weibo, BCI-IV-2a, MunichMI, and Cho2017 datasets have shown increases of 10.2, 7.37, 1.6, 1.3, and 0.75 in their classification accuracies respectively.
On the other hand, the KU, SHU, Murat2018, and HGD datasets did not exhibit noticeable improvements but rather maintained similar performance levels.

Figure \ref{fig:results_NumSub_increasement} also includes a plot illustrating the number of trials for each dataset. By examining the classification accuracy plot in conjunction with the number of trials plot, we can observe that as the number of trials increases, the improvement achieved by the \methodname{} becomes less pronounced. This observation suggests that the proposed framework effectively assists datasets, particularly smaller ones, in leveraging information from datasets with different data formats to enhance their model training and achieve better performance. 

KU, SHU, Murat2018, and HGD datasets exhibit similar performance levels between \methodname{} and baselines. One possible reason for the similar performance is that the number of trials of these datasets is large enough.
Besides, for KU, Murat2018, and HGD datasets, the similar performances may also be due to their high baseline performance, which already exceed 80\%. Thus, they can not benefit significantly from other lower-quality datasets. 
As for the SHU dataset, it may be attributed to the significant disparity between the data distribution of the SHU dataset and the others, as evidenced by its lowest baseline performance among the nine datasets, indicating a large disparity.

\subsubsection{Sensitivity analysis}

\begin{figure}[h!]
  \centering
  \includegraphics[width=0.8\columnwidth]{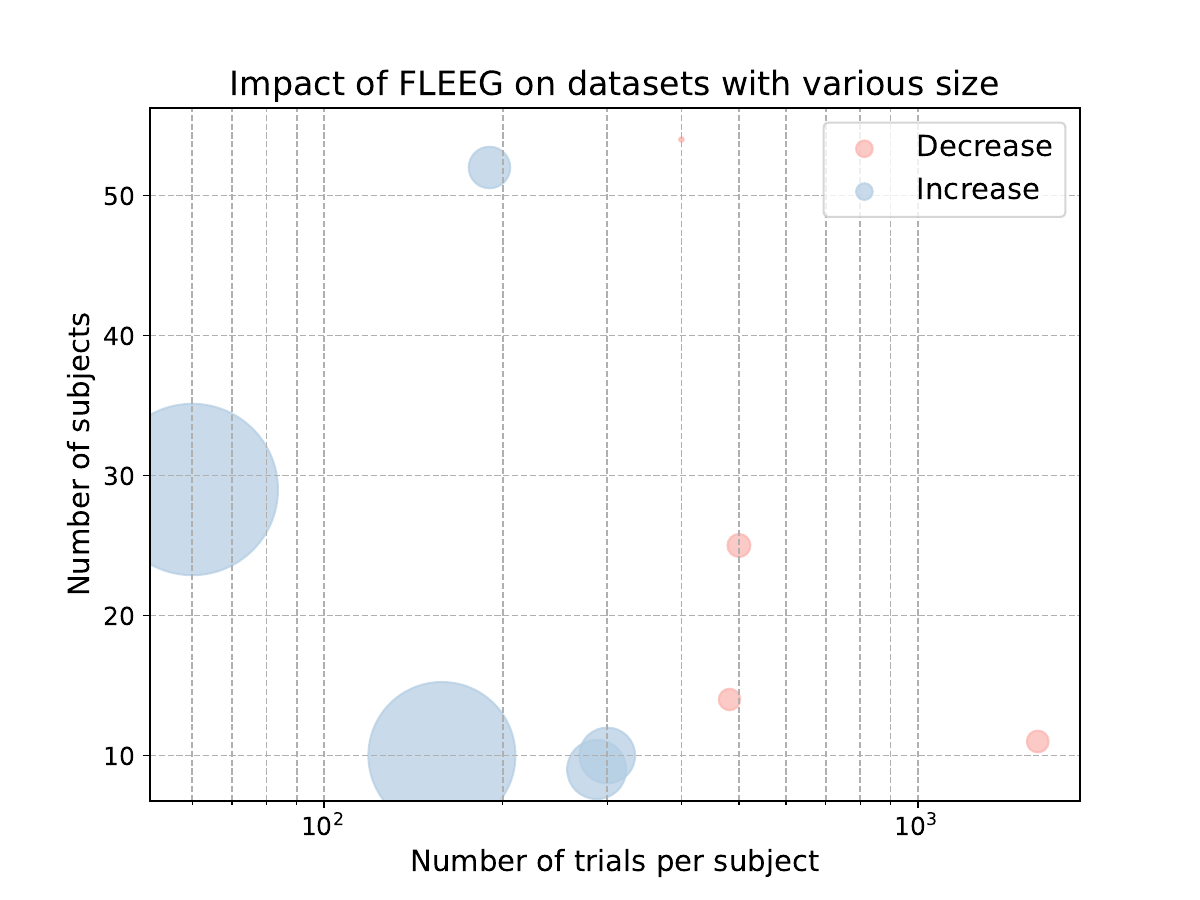}
  \caption{The impact of the framework on the datasets with various numbers of subjects and number of trials per subject. The bubble color indicates the performance of \methodname{} with the blue bubbles illustrating an increase and the red bubbles representing a decrease. The bubble size reflects the absolute values of changes. The number of trials per subject is the primary factor that affects \methodname{} performance, with the number of subjects being a secondary factor.}
  \label{fig:bubblemap}
\end{figure}

Based on the results shown in Figure \ref{fig:results_NumSub_increasement}, it indicates that the performance of \methodname{} is significantly affected by the number of trials. The number of trials is determined by two factors - the number of subjects and the number of trials collected from each subject. To gain a better understanding of how these factors impact the proposed framework, we plot a bubble chart for all datasets in Figure \ref{fig:bubblemap}. The x-axis represents the number of trials per subject, while the y-axis shows the number of subjects in the datasets. The size of each bubble reflects the absolute value of the "\textit{Improvement}" metric, which is defined as $(Acc_{FLEEG}-Acc_{Baseline})/Acc_{Baseline}$. It indicates the performance of \methodname{}: the larger the better for blue bubbles representing positive changes, and the smaller the better for red bubbles representing negative changes. 
The numerical values corresponding to Figure \ref{fig:bubblemap} are provided in Table \ref{tab:Improvement_NumTrialsPerSub} and Table \ref{tab:Improvement_NumSub}. Table \ref{tab:Improvement_NumTrialsPerSub} lists the "\textit{Improvement}" values of all datasets in ascending order of their corresponding numbers of trials per subject. Similarly, Table \ref{tab:Improvement_NumSub} lists the total number of subjects in all datasets in ascending order.

\begin{table}[hbt!]\small
    \begin{subtable}[h]{0.45\textwidth}
        \centering
        \begin{tabular}{c|c|c}
        \hline
        Dataset & \#Trials/Sub & Improvement \\
        \hline \hline
        Shin2017A & 60 & \textbf{16.70\%}\\
        Weibo2014 & 158 & \textbf{12.29\%}\\
        Cho2017 & 190 & \textbf{0.98\%}\\
        BCI-IV-2a & 288 & \textbf{1.99\%}\\
        MunichMI & 300 & \textbf{1.77\%}\\
        KU & 400 & -0.01\%\\
        HGD & 482 & -0.26\%\\
        SHU & 500 & -0.30\% \\
        Murat2018 & 1593 & -0.27\% \\
        \hline
       \end{tabular}
       \caption{Improvements on nine datasets in ascending order of the number of trials per subject.}
       \label{tab:Improvement_NumTrialsPerSub}
    \end{subtable}
    \hfill
    \begin{subtable}[h]{0.45\textwidth}
        \centering
        \begin{tabular}{c|c|c}
        \hline
        Dataset & \#Sub. & Improvement \\
        \hline \hline
        BCI-IV-2a & 9 & \textbf{1.99\%}\\
        Weibo2014 & 10 & \textbf{12.29\%}\\
        MunichMI & 10 & \textbf{1.77\%}\\
        Murat2018 & 11 & -0.27\% \\
        HGD & 14 & -0.26\%\\
        SHU & 25 & -0.30\% \\
        Shin2017A & 29 & \textbf{16.70\%}\\
        Cho2017 & 52 & \textbf{0.98\%}\\
        KU & 54 & -0.01\%\\
        \hline
        \end{tabular}
        \caption{Improvements on nine datasets in ascending order of the number of subjects}
        \label{tab:Improvement_NumSub}
     \end{subtable}
     \caption{Improvements of the \methodname{} with each MI dataset.}
     \label{tab:result_improvement}
\end{table}

In Figure \ref{fig:bubblemap}, bubbles located on the left side of the x-axis are blue while bubbles located on the right side are red. It indicates that datasets with fewer trials per subject can benefit more from the proposed framework. For instance, the Shin2017A dataset, with 60 trials collected for each subject, shows a 16.7\% improvement with \methodname{}. Meanwhile, the Murat2018 dataset, with more than 1500 trials per subject, rarely gains any improvement from the system.
Similarly, for a fixed value on the x-axis, the bubbles at the bottom of the y-axis are larger than the ones at the top. This suggests that datasets with fewer subjects can benefit more from the proposed framework, given a similar number of trials per subject. For example, the Weibo2014 and Cho2017 datasets have around 150-200 trials per subject. But the improvement on the Weibo2014 dataset, with 10 subjects, reaches 12.29\%, compared to only 0.98\% on the Cho2017 dataset, which has 52 subjects.

Comparing these two factors, the number of trials per subject is more important. Even if a dataset involves many subjects, \methodname{} can still improve the performance if the number of trials collected from one subject is small, such as Shin2017A dataset. Therefore, the number of trials per subject is the primary factor, while the number of subjects is a secondary factor in determining the \methodname{} performance.
Thus, the proposed framework can be applied to small datasets to train high-performance models. Additionally, with the help of the \methodname{} framework, the model can achieve good results trained with a small number of trials collected from one subject.





\subsubsection{Interpretability and visualization}
In this section, we employ saliency maps \cite{simonyan2014deep} to visualize the informative regions within the data.
For enhanced visualization, the original saliency map is averaged across the time dimension, resulting in each subject's topological map of the EEG channels.

\begin{figure}[hbt!]
  \centering
  \includegraphics[width=\columnwidth]{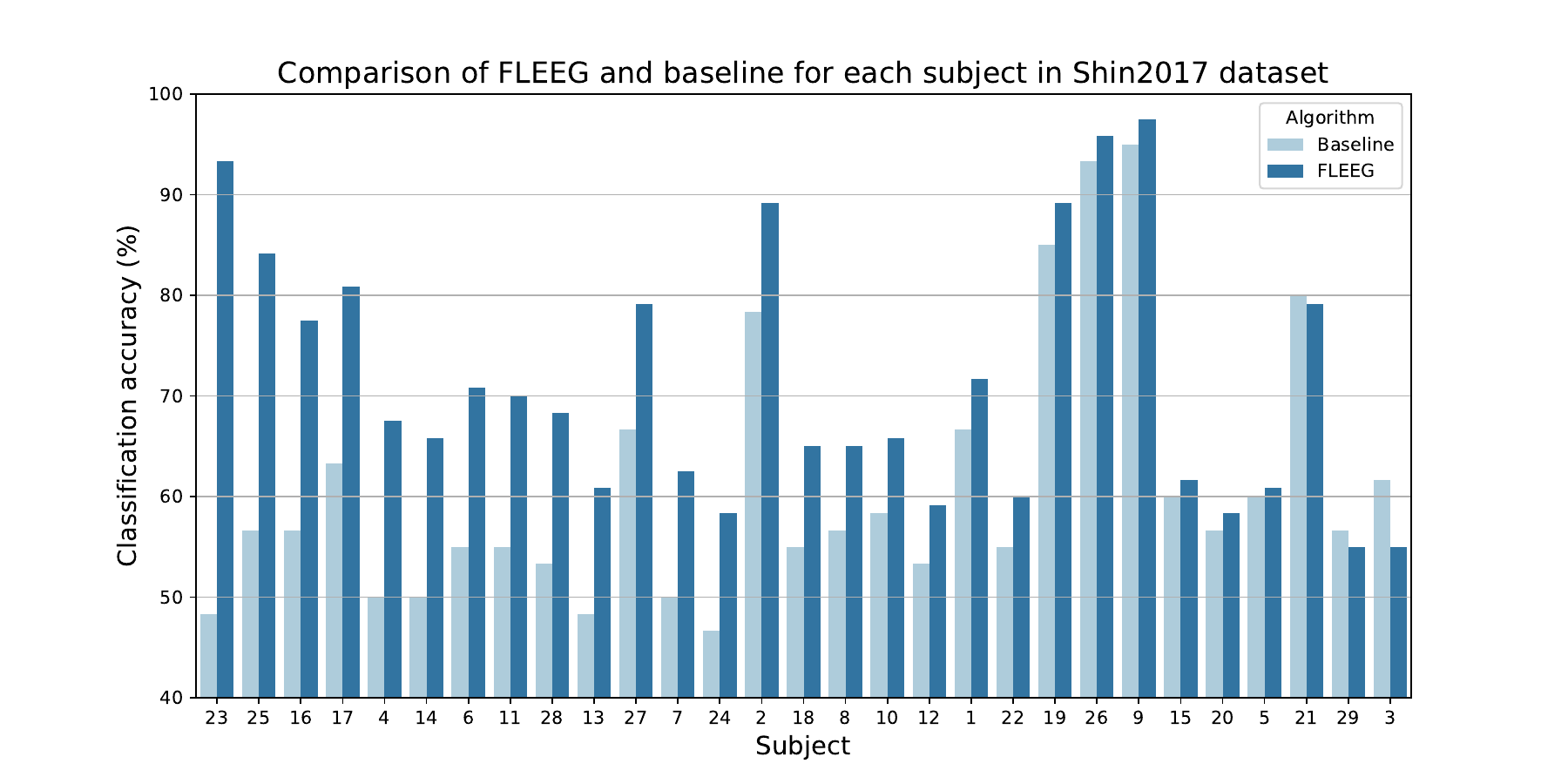}
  \caption{The accuracy comparison of each subject with baseline methods and \methodname{} on the Shin2017 MI EEG datasets.}
  \label{fig:Shin_FLEEGVSDCN}
\end{figure}

\begin{figure} [hbt!]
     \centering
     \begin{subfigure}[b]{0.18\textwidth}
         \centering
         \includegraphics[width=\textwidth]{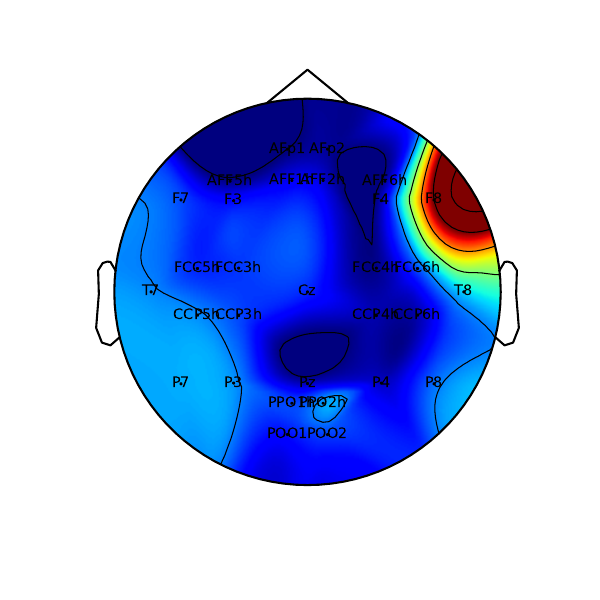}
     \end{subfigure}
     \hfill
     \begin{subfigure}[b]{0.18\textwidth}
         \centering
         \includegraphics[width=\textwidth]{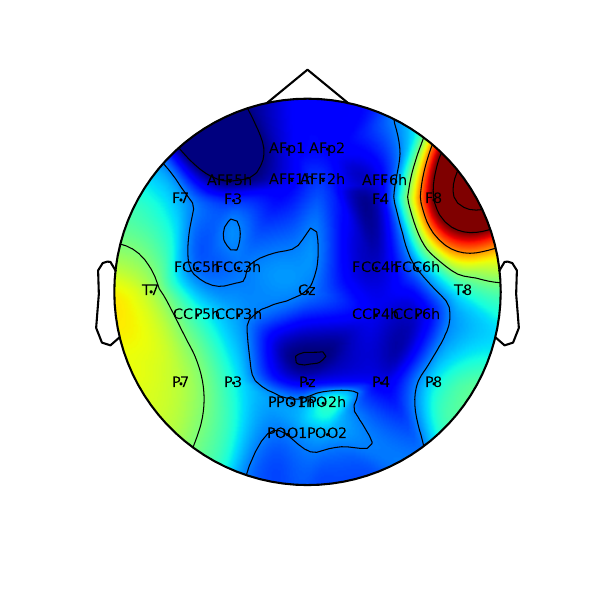}
     \end{subfigure}
     \hfill
     \begin{subfigure}[b]{0.18\textwidth}
         \centering
         \includegraphics[width=\textwidth]{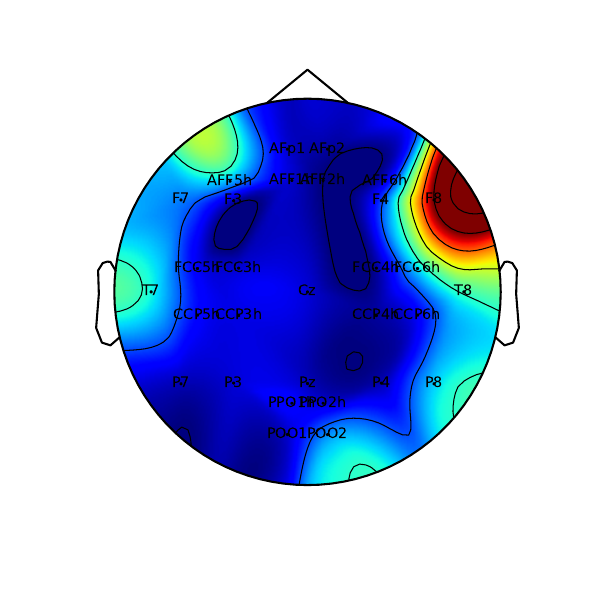}
     \end{subfigure}
     \hfill
     \begin{subfigure}[b]{0.18\textwidth}
         \centering
         \includegraphics[width=\textwidth]{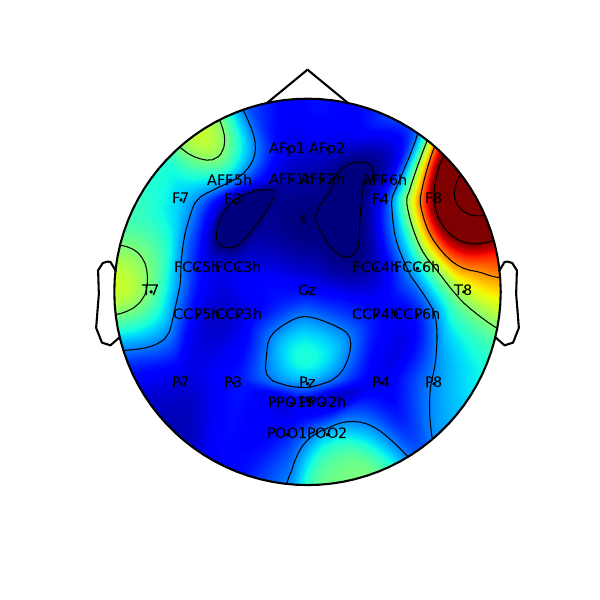}
     \end{subfigure}
     \hfill
     \begin{subfigure}[b]{0.18\textwidth}
         \centering
         \includegraphics[width=\textwidth]{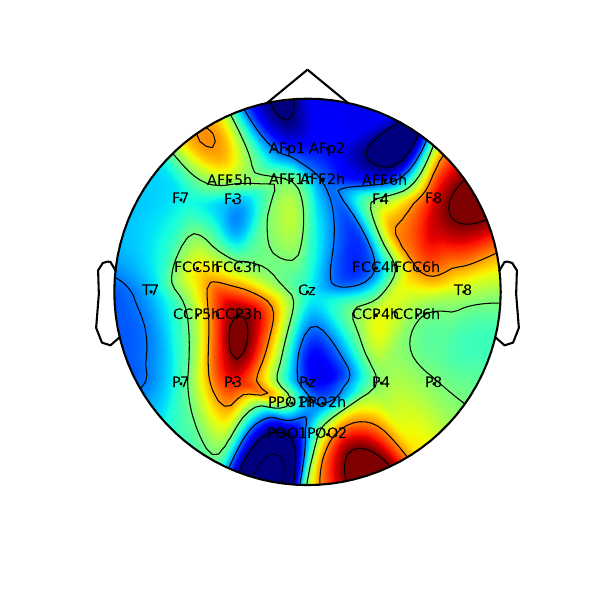}
     \end{subfigure}
     \quad
     \begin{subfigure}[b]{0.18\textwidth}
         \centering
         \includegraphics[width=\textwidth]{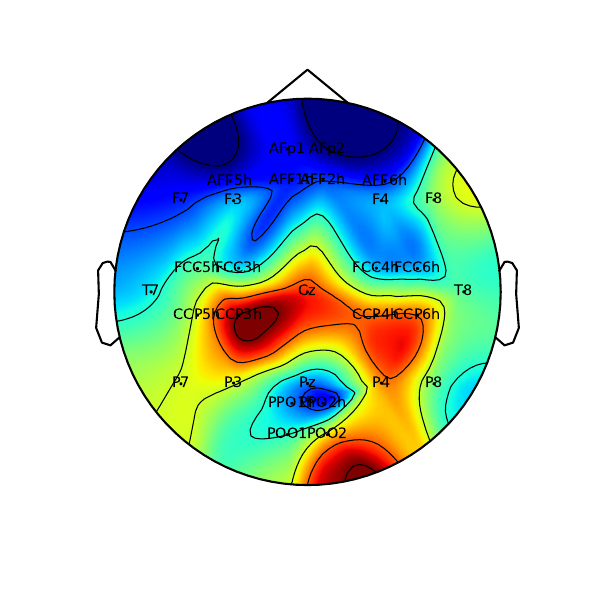}
         \caption{Subject 23}
         \label{fig:Shin_sub23}
     \end{subfigure}
      \hfill
     \begin{subfigure}[b]{0.18\textwidth}
         \centering
         \includegraphics[width=\textwidth]{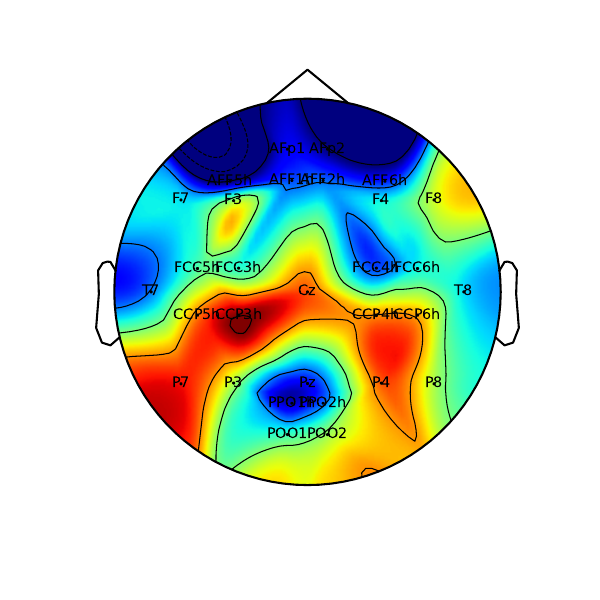}
         \caption{Subject 25}
         \label{fig:Shin_sub25}
     \end{subfigure}
     \hfill
     \begin{subfigure}[b]{0.18\textwidth}
         \centering
         \includegraphics[width=\textwidth]{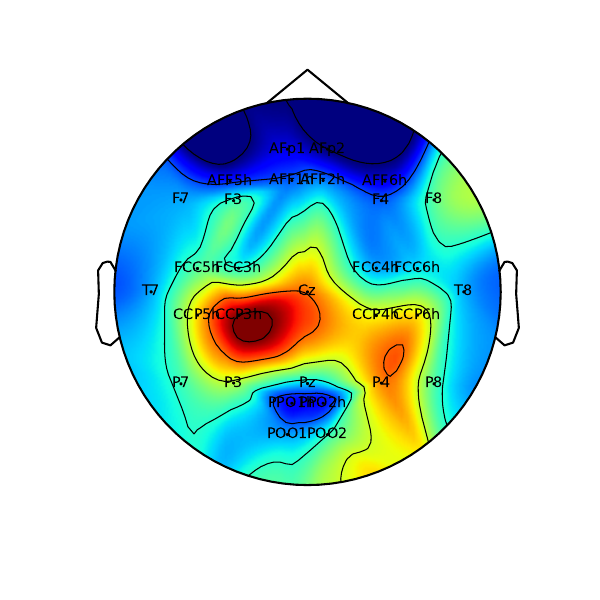}
         \caption{Subject 16}
         \label{fig:Shin_sub16}
     \end{subfigure}
     \hfill
     \begin{subfigure}[b]{0.18\textwidth}
         \centering
         \includegraphics[width=\textwidth]{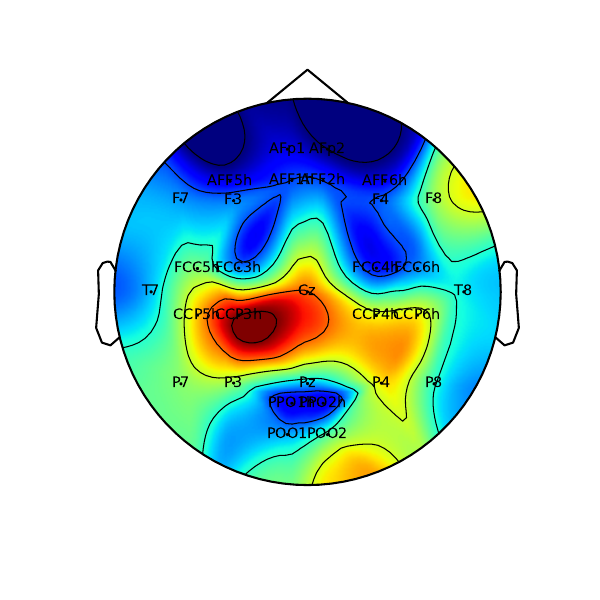}
         \caption{Subject 17}
         \label{fig:Shin_sub17}
     \end{subfigure}
     \hfill
     \begin{subfigure}[b]{0.18\textwidth}
         \centering
         \includegraphics[width=\textwidth]{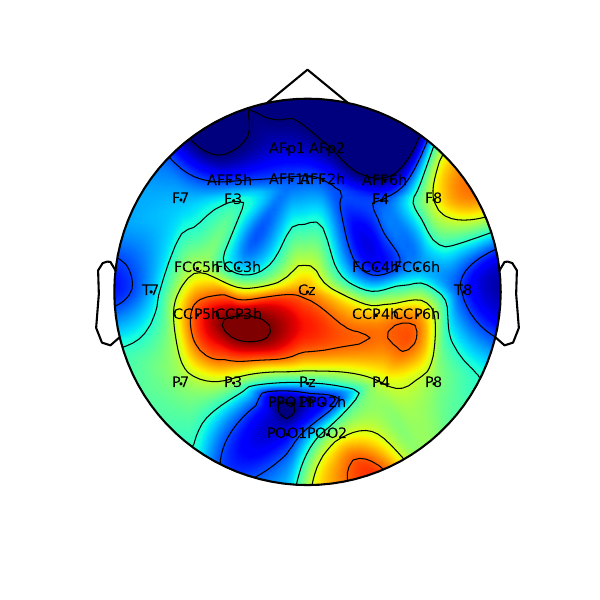}
         \caption{Subject 4}
         \label{fig:Shin_sub4}
     \end{subfigure}
\caption{The saliency maps for the subjects with the top 5 improvements in the Shin2017 dataset. The first row presents the plots for the baseline method and the second row lists the maps for \methodname{}.}
\label{fig:Shin_silencymap_top5}
\end{figure}

\begin{figure}[hbt!]
     \centering
     \begin{subfigure}[b]{0.18\textwidth}
         \centering
         \includegraphics[width=\textwidth]{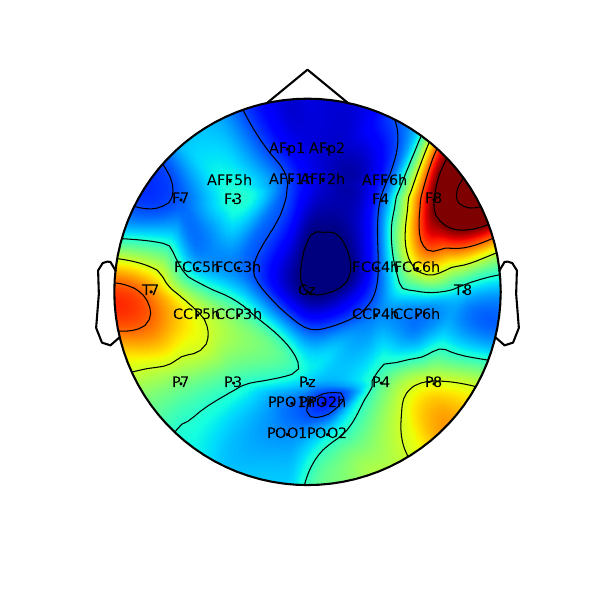}
     \end{subfigure}
     \hfill
     \begin{subfigure}[b]{0.18\textwidth}
         \centering
         \includegraphics[width=\textwidth]{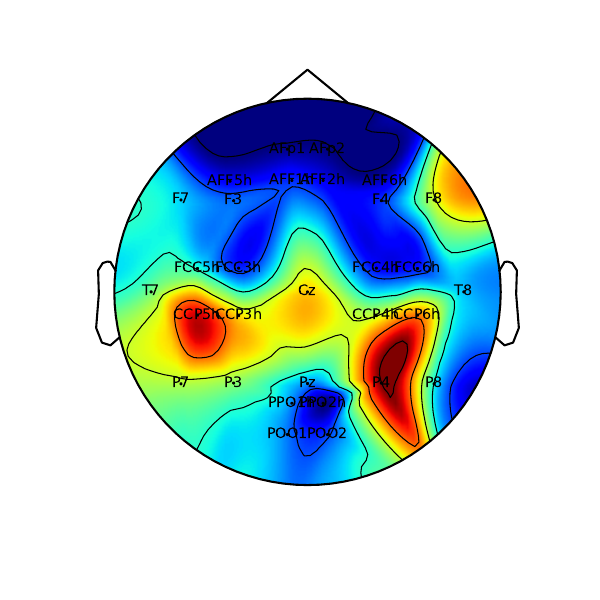}
     \end{subfigure}
     \hfill
     \begin{subfigure}[b]{0.18\textwidth}
         \centering
         \includegraphics[width=\textwidth]{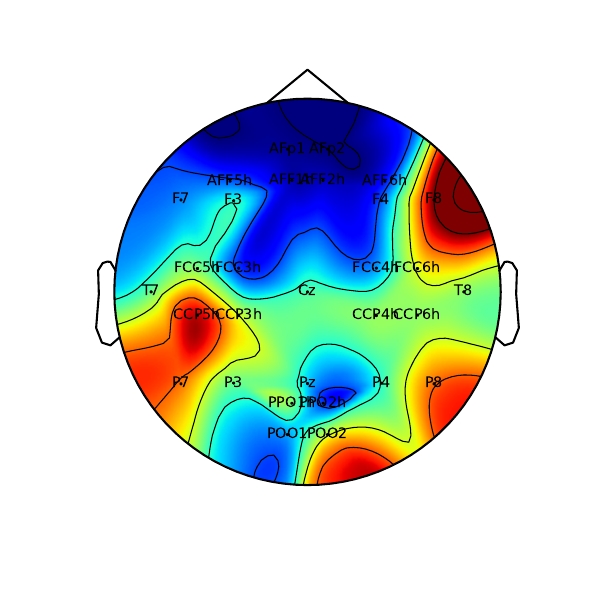}
     \end{subfigure}
     \hfill
     \begin{subfigure}[b]{0.18\textwidth}
         \centering
         \includegraphics[width=\textwidth]{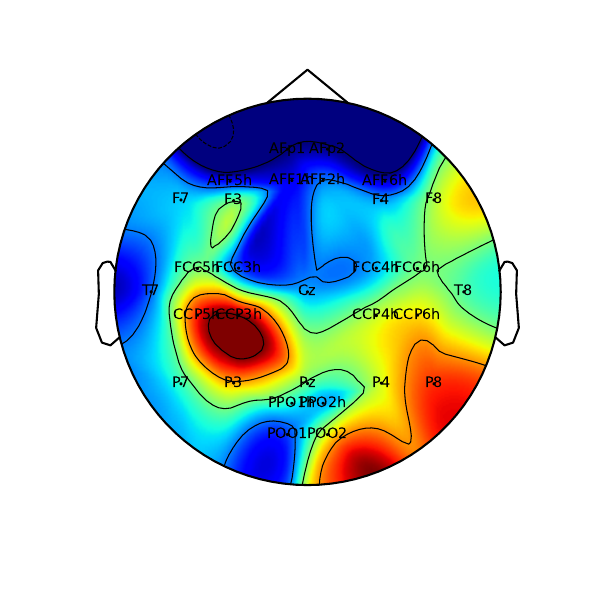}
     \end{subfigure}
     \hfill
     \begin{subfigure}[b]{0.18\textwidth}
         \centering
         \includegraphics[width=\textwidth]{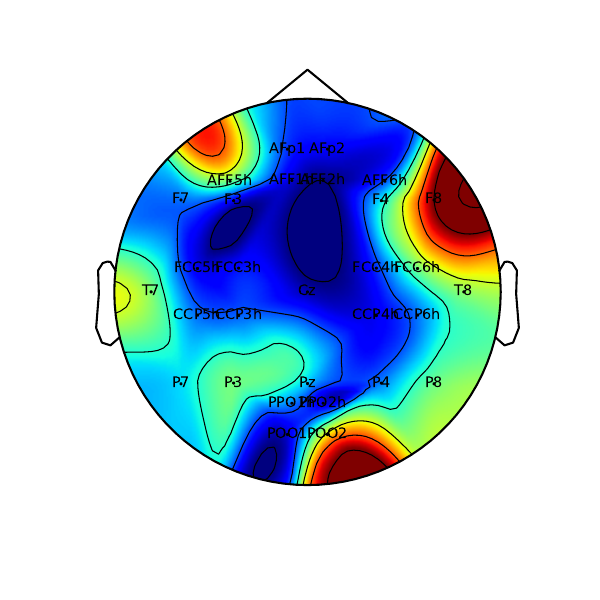}
     \end{subfigure}
     \quad
     \begin{subfigure}[b]{0.18\textwidth}
         \centering
         \includegraphics[width=\textwidth]{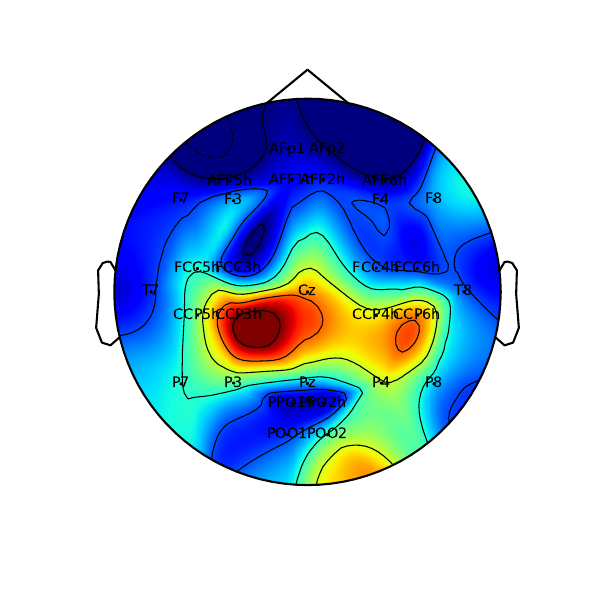}
         \caption{Subject 20}
         \label{fig:Shin_sub20}
     \end{subfigure}
      \hfill
     \begin{subfigure}[b]{0.18\textwidth}
         \centering
         \includegraphics[width=\textwidth]{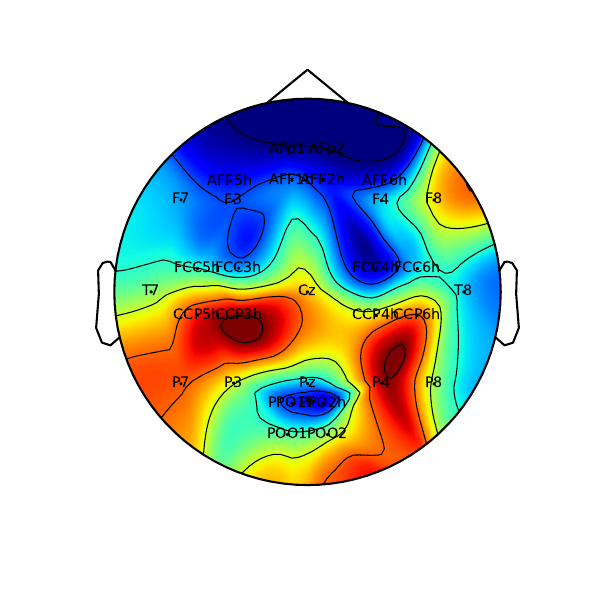}
         \caption{Subject 5}
         \label{fig:Shin_sub5}
     \end{subfigure}
     \hfill
     \begin{subfigure}[b]{0.18\textwidth}
         \centering
         \includegraphics[width=\textwidth]{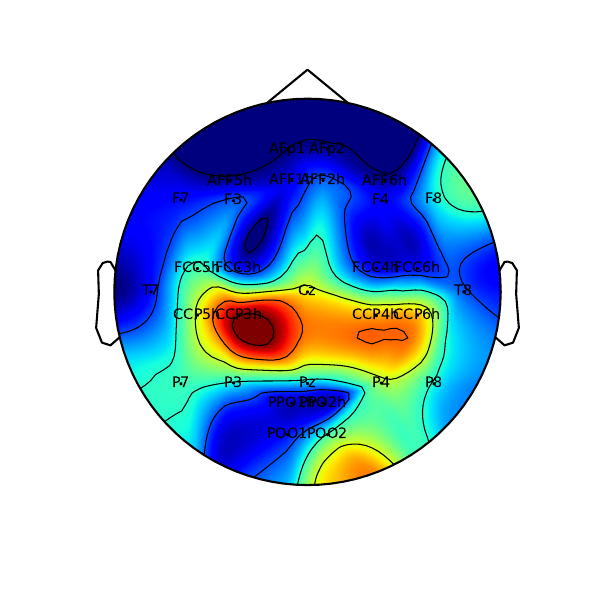}
         \caption{Subject 21}
         \label{fig:Shin_sub21}
     \end{subfigure}
     \hfill
     \begin{subfigure}[b]{0.18\textwidth}
         \centering
         \includegraphics[width=\textwidth]{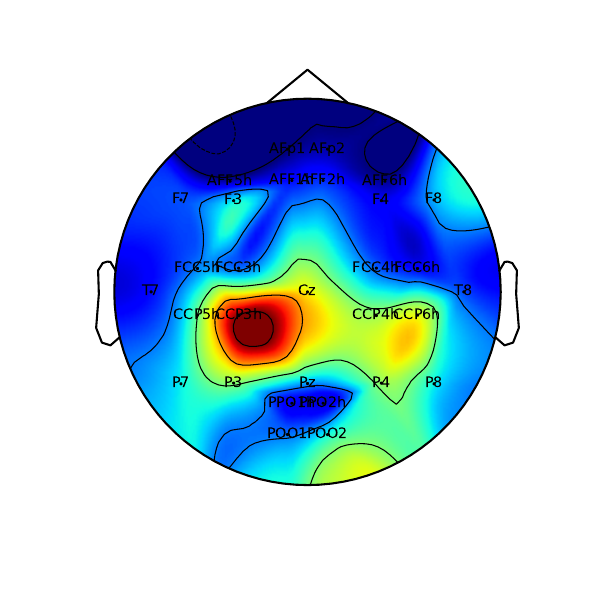}
         \caption{Subject 29}
         \label{fig:Shin_sub29}
     \end{subfigure}
     \hfill
     \begin{subfigure}[b]{0.18\textwidth}
         \centering
         \includegraphics[width=\textwidth]{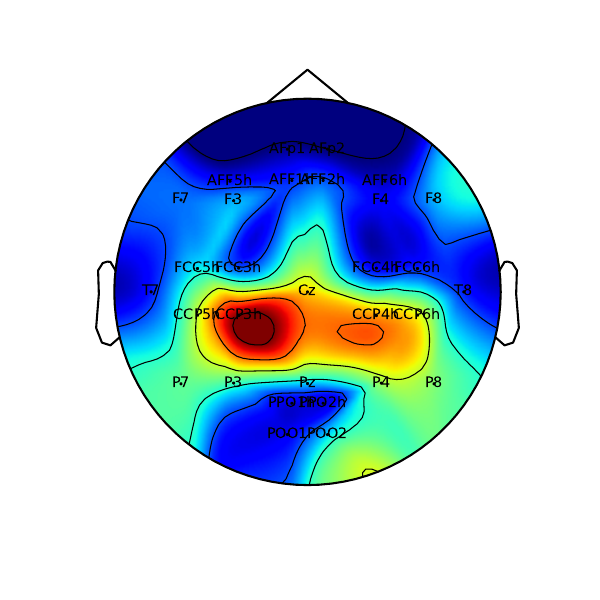}
         \caption{Subject 3}
         \label{fig:Shin_sub3}
     \end{subfigure}
\caption{The saliency maps for the subjects with the bottom 5 improvements in the Shin2017 dataset. The first row presents the plots for the baseline method and the second row lists the maps for \methodname{}.}
\label{fig:Shin_silencymap_bottom5}
\end{figure}

\begin{figure}[hbt!]
  \centering
  \includegraphics[width=\columnwidth]{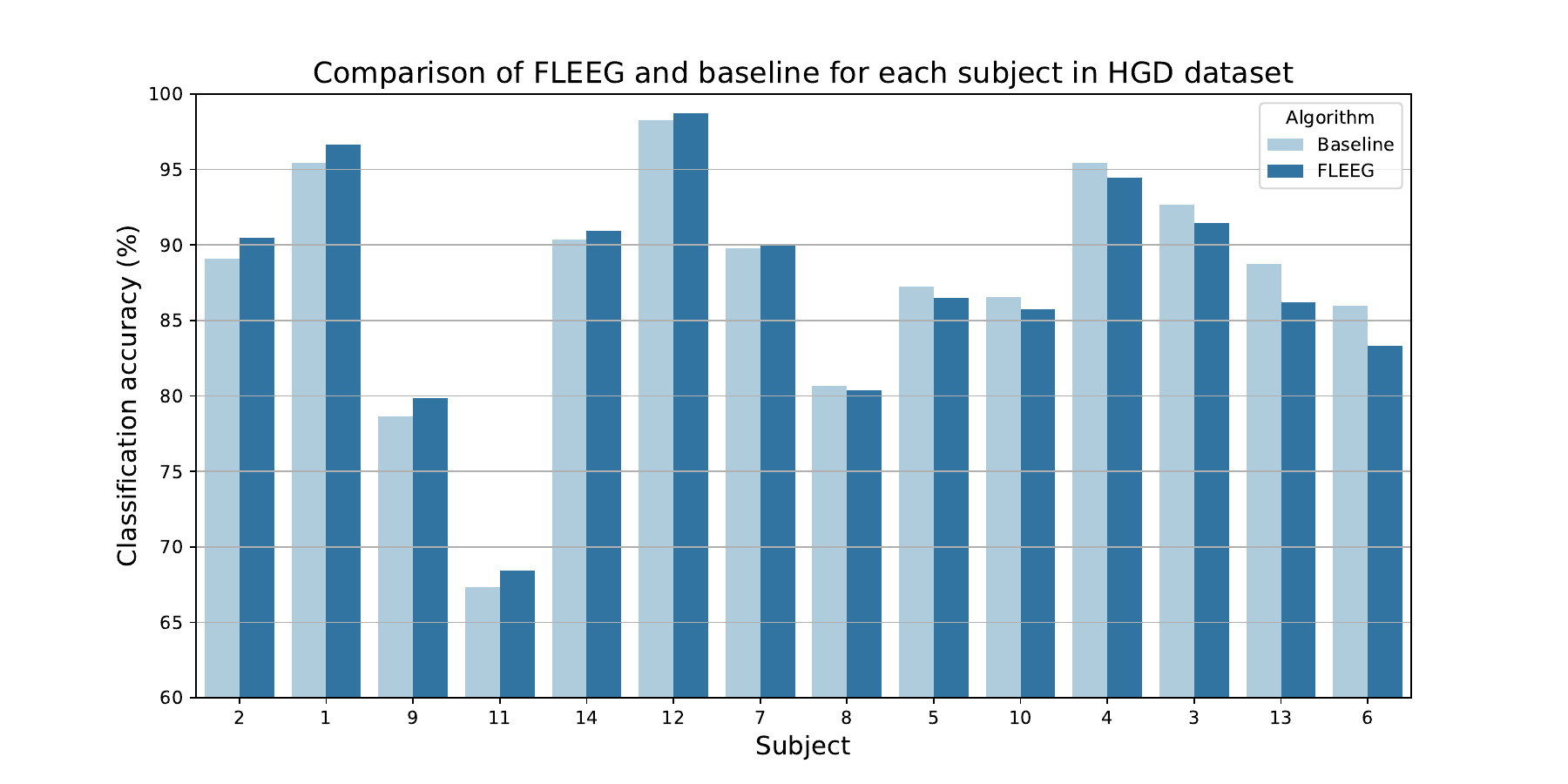}
  \caption{The accuracy comparison of each subject with baseline methods and \methodname{} on the HGD MI EEG datasets.}
  \label{fig:HGD_FLEEGVSDCN}
\end{figure}

\begin{figure} [hbt!]
     \centering
     \begin{subfigure}[b]{0.18\textwidth}
         \centering
         \includegraphics[width=\textwidth]{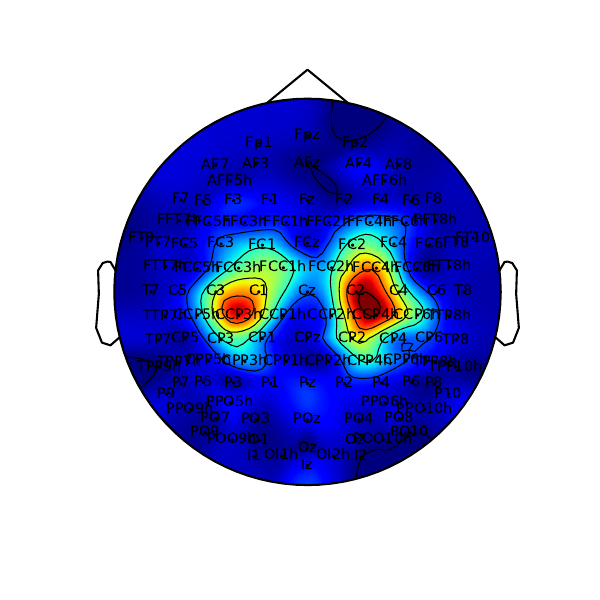}
     \end{subfigure}
     \hfill
     \begin{subfigure}[b]{0.18\textwidth}
         \centering
         \includegraphics[width=\textwidth]{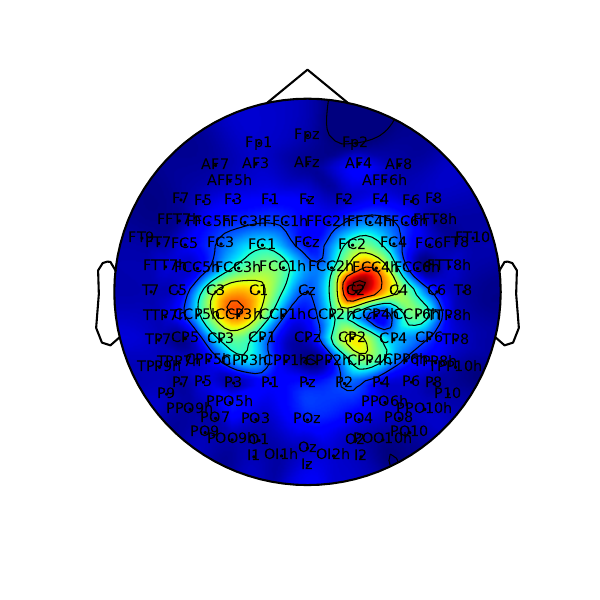}
     \end{subfigure}
     \hfill
     \begin{subfigure}[b]{0.18\textwidth}
         \centering
         \includegraphics[width=\textwidth]{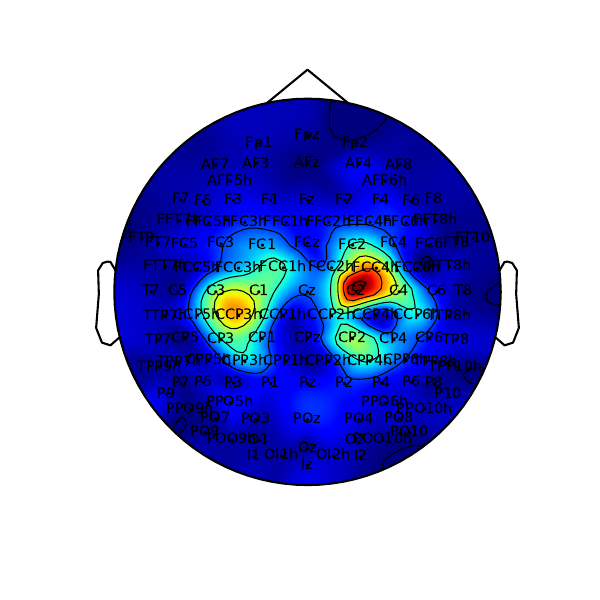}
     \end{subfigure}
     \hfill
     \begin{subfigure}[b]{0.18\textwidth}
         \centering
         \includegraphics[width=\textwidth]{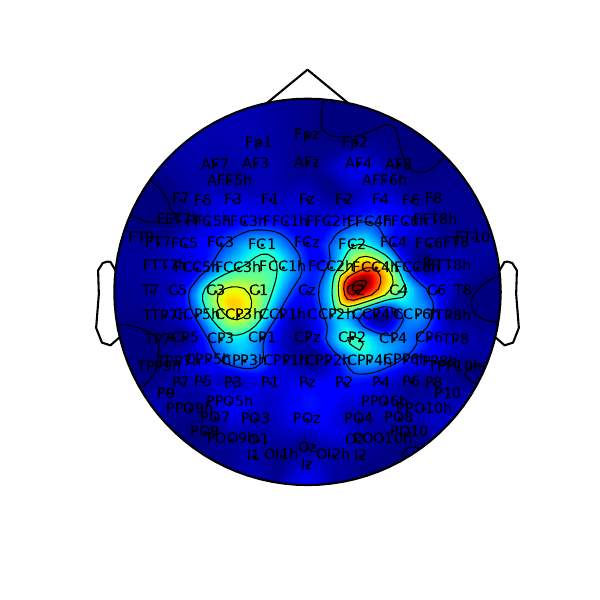}
     \end{subfigure}
     \hfill
     \begin{subfigure}[b]{0.18\textwidth}
         \centering
         \includegraphics[width=\textwidth]{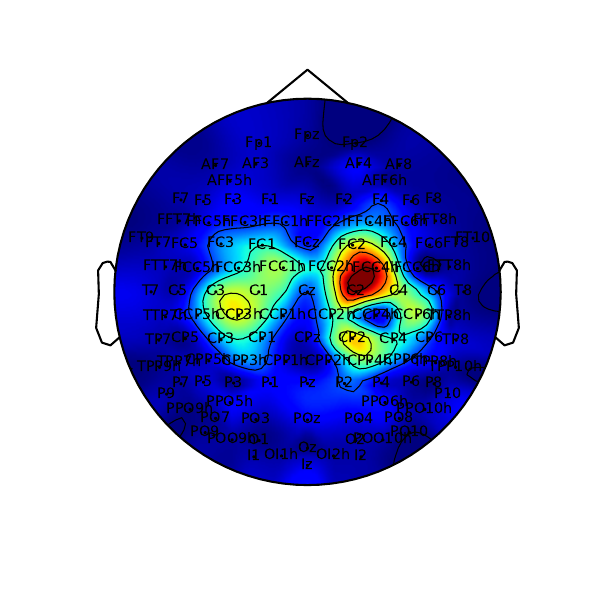}
     \end{subfigure}
     \quad
     \begin{subfigure}[b]{0.18\textwidth}
         \centering
         \includegraphics[width=\textwidth]{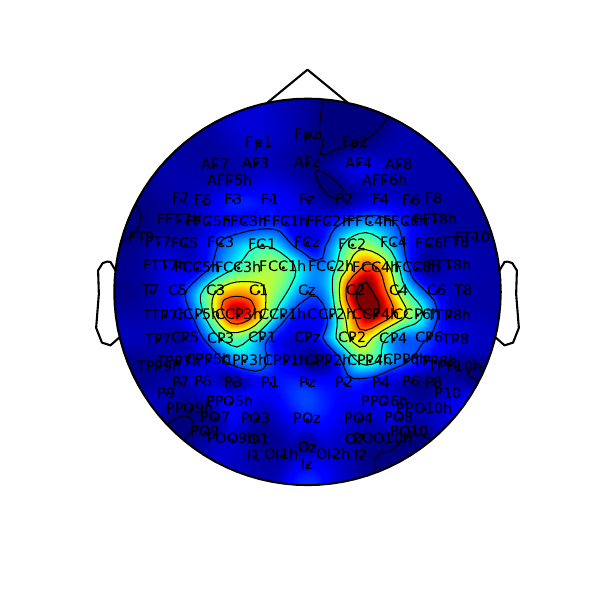}
         \caption{Subject 2}
         \label{fig:HGE_sub2}
     \end{subfigure}
      \hfill
     \begin{subfigure}[b]{0.18\textwidth}
         \centering
         \includegraphics[width=\textwidth]{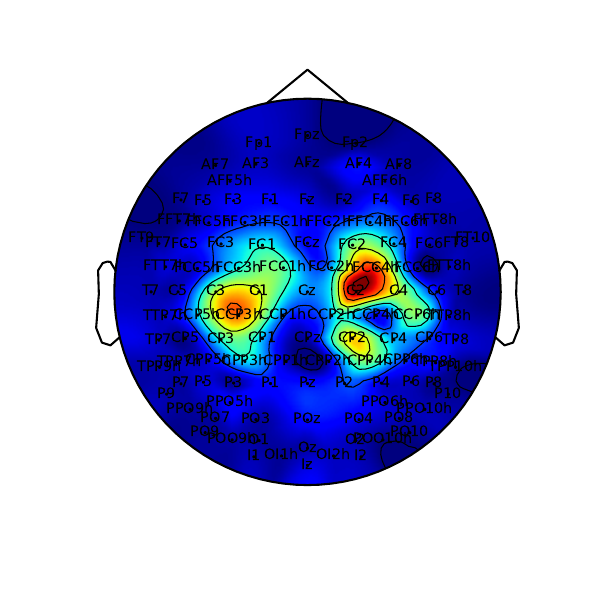}
         \caption{Subject 1}
         \label{fig:HGD_sub1}
     \end{subfigure}
     \hfill
     \begin{subfigure}[b]{0.18\textwidth}
         \centering
         \includegraphics[width=\textwidth]{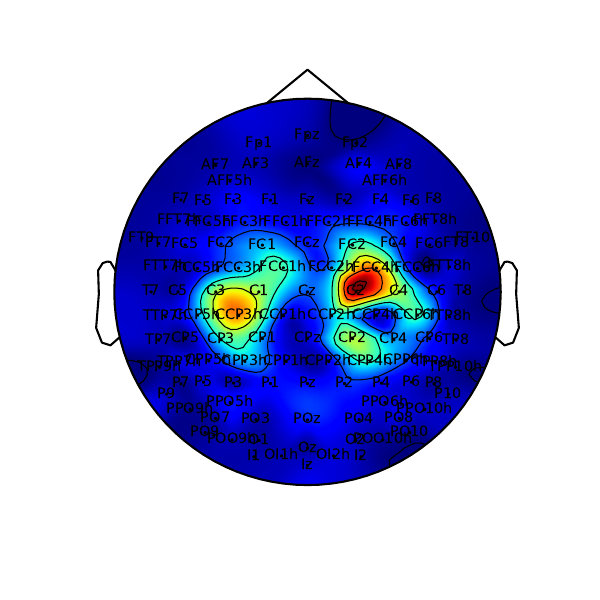}
         \caption{Subject 9}
         \label{fig:HGD_sub9}
     \end{subfigure}
     \hfill
     \begin{subfigure}[b]{0.18\textwidth}
         \centering
         \includegraphics[width=\textwidth]{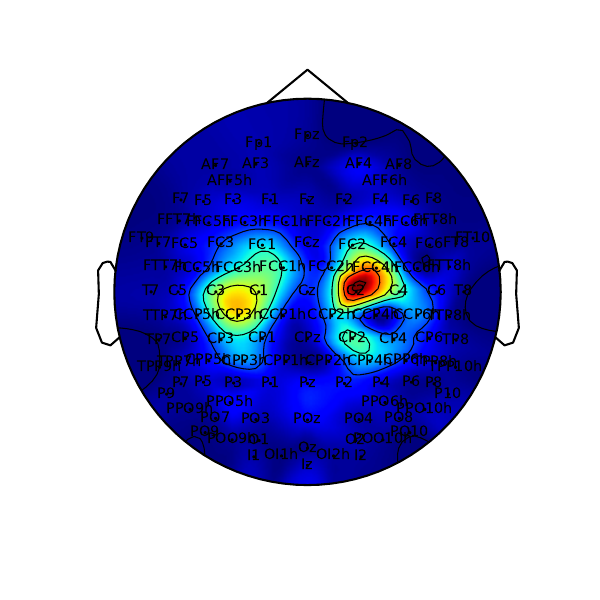}
         \caption{Subject 11}
         \label{fig:HGD_sub11}
     \end{subfigure}
     \hfill
     \begin{subfigure}[b]{0.18\textwidth}
         \centering
         \includegraphics[width=\textwidth]{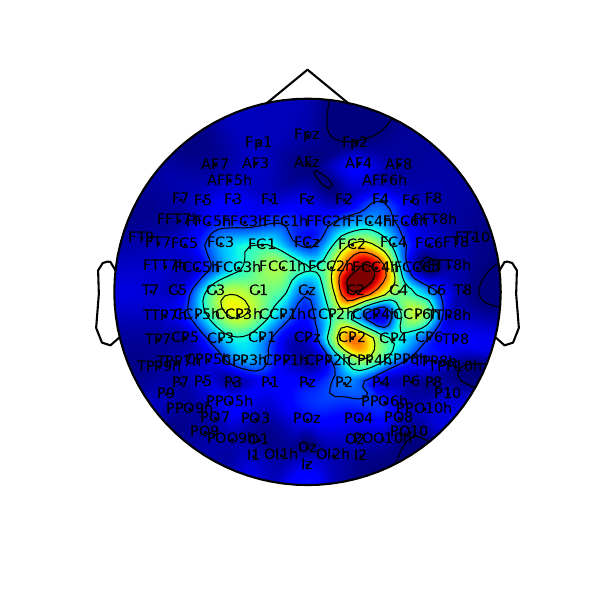}
         \caption{Subject 14}
         \label{fig:HGD_sub14}
     \end{subfigure}
\caption{The saliency maps for the subjects with the top 5 improvements in the HGD dataset. The first row presents the plots for the baseline method and the second row lists the maps for \methodname{}.}
\label{fig:HGD_silencymap_top5}
\end{figure}

\begin{figure}[hbt!]
     \centering
     \begin{subfigure}[b]{0.18\textwidth}
         \centering
         \includegraphics[width=\textwidth]{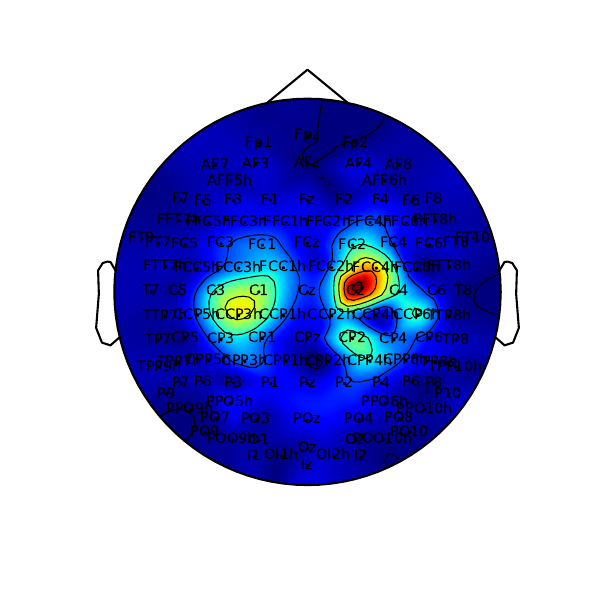}
     \end{subfigure}
     \hfill
     \begin{subfigure}[b]{0.18\textwidth}
         \centering
         \includegraphics[width=\textwidth]{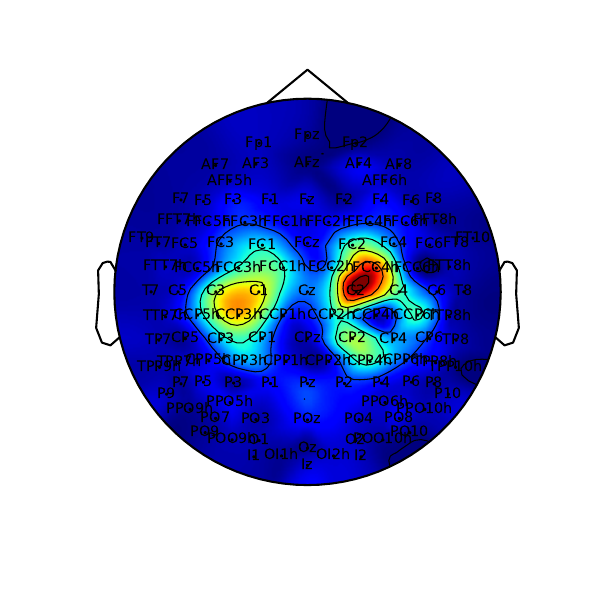}
     \end{subfigure}
     \hfill
     \begin{subfigure}[b]{0.18\textwidth}
         \centering
         \includegraphics[width=\textwidth]{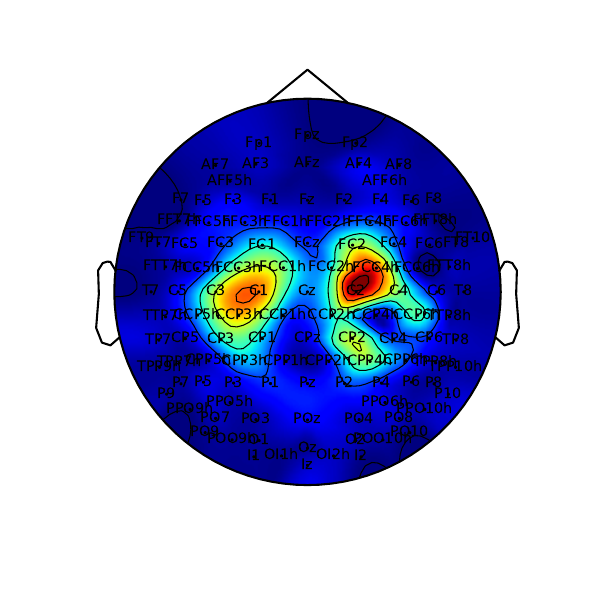}
     \end{subfigure}
     \hfill
     \begin{subfigure}[b]{0.18\textwidth}
         \centering
         \includegraphics[width=\textwidth]{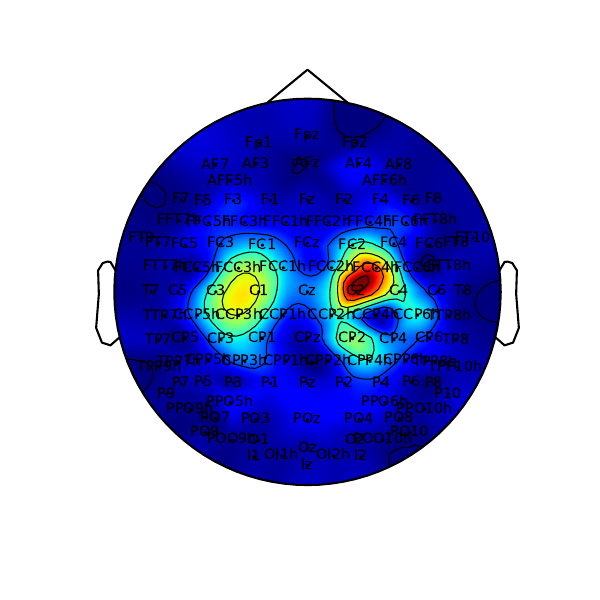}
     \end{subfigure}
     \hfill
     \begin{subfigure}[b]{0.18\textwidth}
         \centering
         \includegraphics[width=\textwidth]{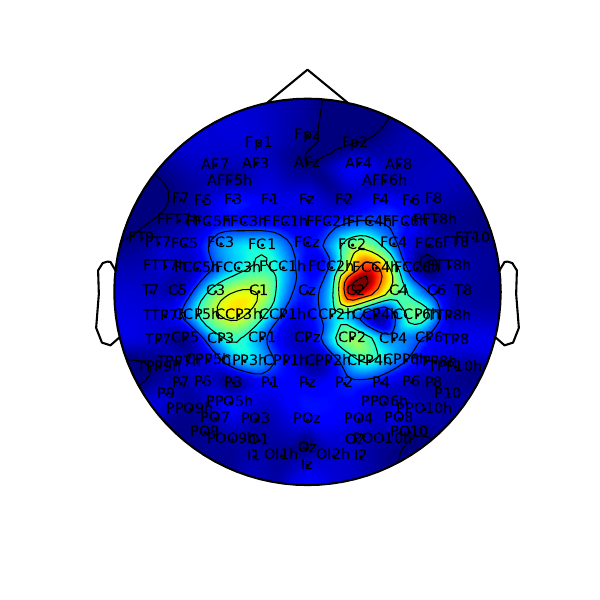}
     \end{subfigure}
     \quad
     \begin{subfigure}[b]{0.18\textwidth}
         \centering
         \includegraphics[width=\textwidth]{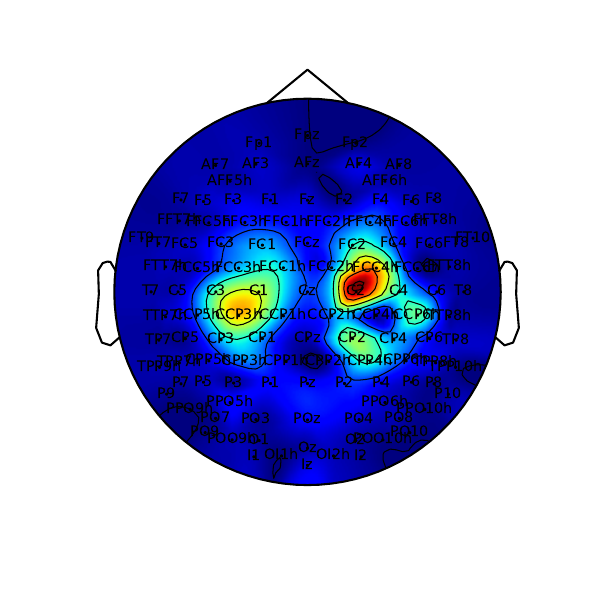}
         \caption{Subject 10}
         \label{fig:HGD_sub10}
     \end{subfigure}
      \hfill
     \begin{subfigure}[b]{0.18\textwidth}
         \centering
         \includegraphics[width=\textwidth]{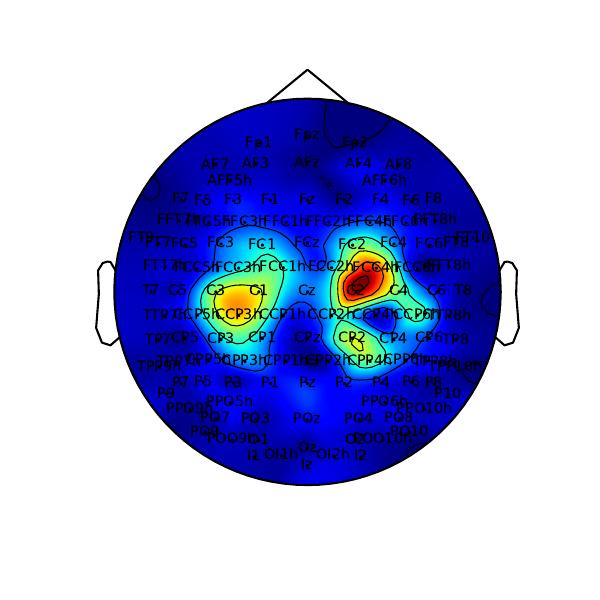}
         \caption{Subject 4}
         \label{fig:HGD_sub4}
     \end{subfigure}
     \hfill
     \begin{subfigure}[b]{0.18\textwidth}
         \centering
         \includegraphics[width=\textwidth]{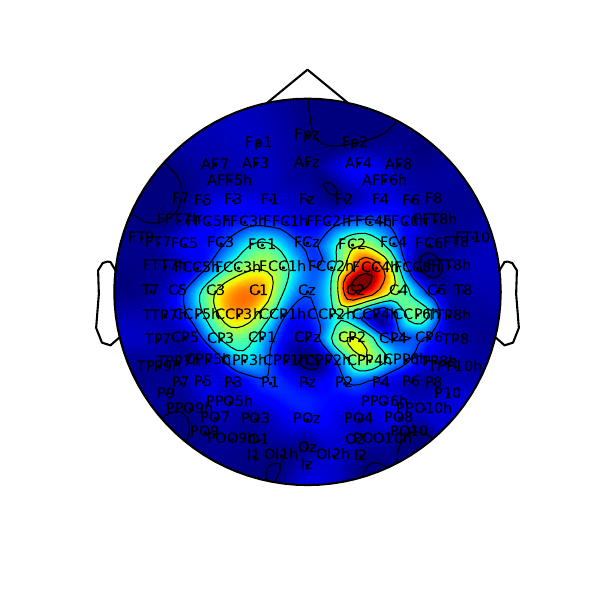}
         \caption{Subject 3}
         \label{fig:HGD_sub3}
     \end{subfigure}
     \hfill
     \begin{subfigure}[b]{0.18\textwidth}
         \centering
         \includegraphics[width=\textwidth]{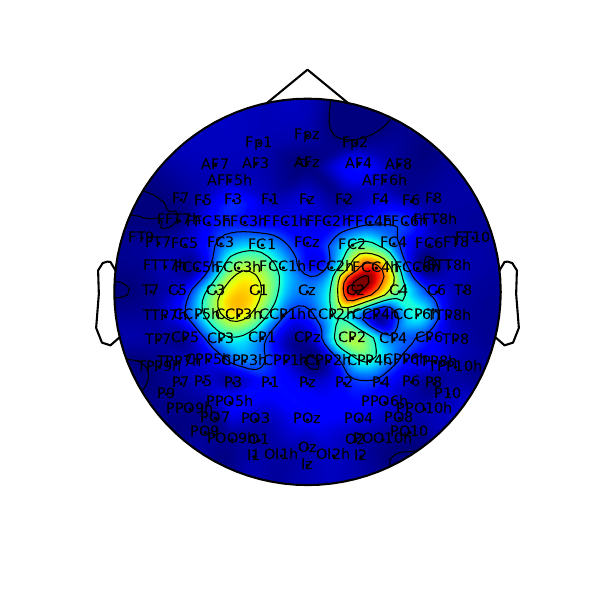}
         \caption{Subject 13}
         \label{fig:HGD_sub13}
     \end{subfigure}
     \hfill
     \begin{subfigure}[b]{0.18\textwidth}
         \centering
         \includegraphics[width=\textwidth]{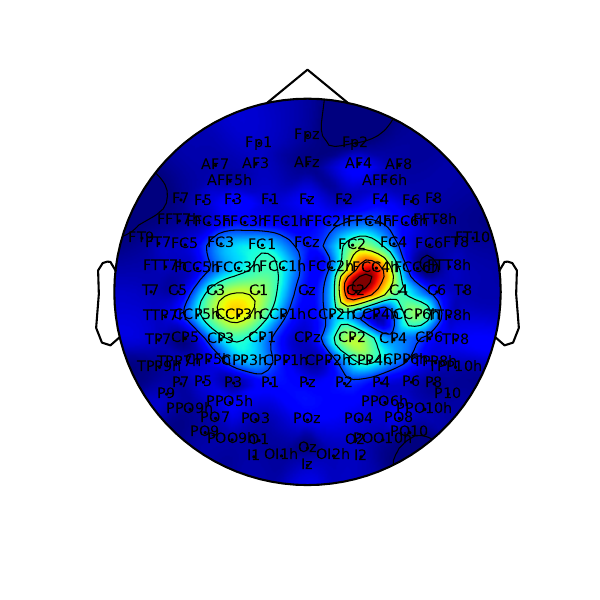}
         \caption{Subject 6}
         \label{fig:HGD_sub6}
     \end{subfigure}
\caption{The saliency maps for the subjects with the bottom 5 improvements in the HGD dataset. The first row presents the plots for the baseline method and the second row lists the maps for \methodname{}.}
\label{fig:HGD_silencymap_bottom5}
\end{figure}

We plot the saliency maps of the Shin2017A and HGD datasets. Shin2017A gains the largest improvement from the framework, meanwhile, HGD has the largest decrease. 
We also plot the individual accuracy comparison for each subject in these two datasets.
For Shin2017A dataset, the accuracy comparison for each subject is presented in Figure \ref{fig:Shin_FLEEGVSDCN} with a descending sequence of accuracy improvement. 
Due to the space limitation, we only plot the saliency map of subjects with the top 5 improvements and bottom 5 improvements in Figure \ref{fig:Shin_silencymap_top5} and Figure \ref{fig:Shin_silencymap_bottom5}, correspondingly.
For HGD dataset, the accuracy comparison for each subject is presented in \ref{fig:HGD_FLEEGVSDCN}, sorted with a descending sequence of accuracy improvement. 
Due to the space limitation, we also plot the saliency map of the top 5 and bottom 5 subjects in Figure \ref{fig:HGD_silencymap_top5} and Figure \ref{fig:HGD_silencymap_bottom5}, correspondingly.

Compared with the baseline, the proposed framework can stably catch the features from the most informative areas related to the motor cortex regions in the brain, even for small datasets. According to Figure \ref{fig:Shin_silencymap_top5}, F3 contributes more to the predictions of the baseline method, whereas, CCP5h, CCP3h, Cz, CCP4h, and CCP6h provide more information to \methodname{}. This indicates that \methodname{} learns neurophysiologically meaningful features from the EEG signals originating from motor cortex regions \cite{PFURTSCHELLER199765}. Although the improvements are relatively lower for the bottom 5 subjects shown in Figure \ref{fig:Shin_silencymap_bottom5}, \methodname{} stably learns from motor cortex regions ( CCP5h and CCP3h ) compared to the baseline method that focuses on non-motor areas for some subjects, i.e., subject 20 and subject 3. 
As the number of samples increases, both \methodname{} and the baseline method concentrate more on EEG from the motor-related areas. As shown in Figure \ref{fig:HGD_silencymap_top5} and Figure \ref{fig:HGD_silencymap_bottom5}, both methods focus on CCP3h, CCP4h, and C2 which are located in the motor area of the brain when they are trained on the HGD dataset which has more data samples than Shin2017.


\section{CONCLUSIONS AND FUTURE WORK}
\label{sec:conclusion}

In this work, we proposed a new learning paradigm for BCI to train the high-performance EEG decoding model with multiple datasets. We designed a hierarchical personalized federated-learning-based framework \methodname{} to solve the device-heterogeneity issue among multiple EEG datasets, enabling knowledge sharing between datasets.
The proposed framework has been evaluated with nine real MI datasets and obtained promising results with reasonable interpretations.
This framework overcomes the challenge of insufficient data for model training in BCI. Thus, small datasets can train better models by making use of the knowledge from other datasets with the help of \methodname{}. 

In the future, it is interesting to apply the proposed framework to more complex situations where the datasets have various protocols and tasks or improve the local module with more powerful feature encoders.

\begin{sidewaystable}[h]\small
    \centering
    \begin{tabular}{l|l|ll|lll|lll|lll|l}
    \hline
    \multicolumn{1}{c|}{\multirow{2}{*}{Dataset}} &
      \multirow{2}{*}{\makecell[l]{Input\\ size}} &
      \multicolumn{2}{c|}{Conv-temp} &
      \multicolumn{3}{c|}{Conv-spatial} &
      \multicolumn{3}{c|}{Conv-pool} &
      \multicolumn{3}{c|}{Conv-pool} &
      \multirow{2}{*}{\makecell[l]{Output \\size}} \\ \cline{3-13}
    \multicolumn{1}{c|}{} &
       &
      \makecell[l]{num. \\ of \\ker.} &
      \makecell[l]{ker. \\size} &
      \makecell[l]{num. \\ of \\ker.} &
      \makecell[l]{ker. \\size} &
      \makecell[l]{pool. \\ker. \\size} &
      \makecell[l]{num. \\ of \\ker.} &
      \makecell[l]{ker. \\size} &
      \makecell[l]{pool. \\ker. \\size} &
      \makecell[l]{num. \\ of \\ker.} &
      \makecell[l]{ker. \\size} &
      \makecell[l]{pool. \\ker. \\size} &
        \\ \hline
        KU        & {[}1,62,1000{]}  & 25 & (1,10) & 25 & (62,1)  & (1,3) & 50 & (1,10) & (1,3) & 100 & (1,10) & (1,3) & {[}100,1,32{]} \\
        SHU       & {[}1,32,1000{]}  & 25 & (1,10) & 25 & (32,1)  & (1,3) & 50 & (1,10) & (1,3) & 100 & (1,10) & (1,3) & {[}100,1,32{]} \\
        Shin2017  & {[}1,30,2000{]}  & 25 & (1,8)  & 25 & (30,1)  & (1,5) & 50 & (1,8)  & (1,4) & 100 & (1,8)  & (1,3) & {[}100,1,30{]} \\
        BCI-IV-2a & {[}1,22,1000{]}  & 25 & (1,10) & 25 & (22,1)  & (1,3) & 50 & (1,10) & (1,3) & 100 & (1,10) & (1,3) & {[}100,1,32{]} \\
        Weibo2014 & {[}1,60,800{]}   & 25 & (1,8)  & 25 & (60,1)  & (1,2) & 50 & (1,8)  & (1,3) & 100 & (1,8)  & (1,3) & {[}100,1,30{]} \\
        MunichMI  & {[}1,128,3500{]} & 25 & (1,10) & 25 & (128,1) & (1,4) & 50 & (1,10) & (1,4) & 100 & (1,10) & (1,3) & {[}100,1,32{]} \\
        HGD       & {[}1,128,2000{]} & 25 & (1,20) & 25 & (128,1) & (1,6) & 50 & (1,20) & (1,3) & 100 & (1,10) & (1,3) & {[}100,1,31{]} \\
        Cho2017   & {[}1,64,1536{]}  & 25 & (1,22) & 25 & (64,1)  & (1,4) & 50 & (1,22) & (1,3) & 100 & (1,22) & (1,3) & {[}100,1,32{]} \\
        Murat2018 & {[}1,22,200{]}   & 25 & (1,6)  & 25 & (22,1)  & (1,1) & 50 & (1,6)  & (1,2) & 100 & (1,6)  & (1,3) & {[}100,1,30{]} \\ \hline             
    \end{tabular}
    \caption{The network structure of the local modules for each dataset.}
 \label{tab:local_module_network_settings}
\end{sidewaystable}

\begin{table}[h]\footnotesize
\centering
\begin{tabular}{l|l|lll|ll|l|l}
\hline
\multicolumn{1}{c|}{\multirow{2}{*}{Dataset}} &
  \multirow{2}{*}{Input size} &
  \multicolumn{3}{c|}{Conv-pool} &
  \multicolumn{2}{c|}{Conv-pool} &
  \multicolumn{1}{c|}{\multirow{2}{*}{Softmax}} &
  \multicolumn{1}{c}{\multirow{2}{*}{\makecell[l]{Output\\ size}}} \\ \cline{3-7}
\multicolumn{1}{c|}{} &
   &
  \multicolumn{1}{c}{\makecell[l]{num. \\ of\\ ker.}} &
  \multicolumn{1}{c}{\makecell[l]{ker. \\size}} &
  \multicolumn{1}{c|}{\makecell[l]{pool. \\ ker.\\ size}} &
  \multicolumn{1}{c}{\makecell[l]{num. \\ of\\ ker.}} &
  \multicolumn{1}{c|}{\makecell[l]{num. \\ of\\ ker.}} &
  \multicolumn{1}{c|}{} &
  \multicolumn{1}{c}{} \\ \hline
KU &
  {[}100,1,32{]} &
  \multirow{9}{*}{200} &
  \multirow{9}{*}{(1,10)} &
  \multirow{9}{*}{(1,3)} &
  \multirow{9}{*}{2} &
  \multirow{9}{*}{(1,7)} &
  \multirow{9}{*}{} &
  \multirow{9}{*}{{[}2,1,1{]}} \\
SHU       & {[}100,1,32{]} &  &  &  &  &  &  &  \\
Shin2017  & {[}100,1,30{]} &  &  &  &  &  &  &  \\
BCI-IV-2a & {[}100,1,32{]} &  &  &  &  &  &  &  \\
Weibo2014 & {[}100,1,30{]} &  &  &  &  &  &  &  \\
MunichMI  & {[}100,1,32{]} &  &  &  &  &  &  &  \\
HGD       & {[}100,1,31{]} &  &  &  &  &  &  &  \\
Cho2017   & {[}100,1,32{]} &  &  &  &  &  &  &  \\
Murat2018 & {[}100,1,30{]} &  &  &  &  &  &  &  \\ \hline
\end{tabular}
\caption{The network structure of the global module in the proposed algorithm.}
\label{tab:global_module_network_settings}
\end{table}

\section*{Acknowledgments}
This research/project is supported by the RIE 2020 Advanced Manufacturing and Engineering (AME) Programmatic Fund (No. A20G8b0102), Singapore; the National Research Foundation, Singapore and DSO National Laboratories under the AI Singapore Programme (AISG Award No: AISG2-RP-2020-019);  Nanyang Technological University, Nanyang Assistant Professorship (NAP); and Future Communications Research \& Development Programme (FCP-NTU-RG-2021-014).










\bibliography{ref}
\bibliographystyle{elsarticle-num-names}

\end{document}